\documentclass[aps,pra,twocolumn,groupedaddress,nofootinbib]{revtex4-1}
\usepackage{amsmath}
\usepackage{amssymb}
\usepackage{graphicx}
\usepackage{mathrsfs}
\usepackage{ntheorem}
\usepackage{mathtools}
\usepackage{enumerate}
\usepackage{enumitem}
\usepackage[T1]{fontenc}
\usepackage{physics}
\usepackage{subfigure}
\usepackage{overpic}
\usepackage{epstopdf}
\usepackage{bm}
\usepackage{color,soul}
\usepackage[bookmarks=false]{hyperref}
\usepackage{xcolor}
\hypersetup{colorlinks=true,citecolor=blue,
linkcolor=blue,urlcolor=blue,pdfstartview=FitH,
bookmarksopen=true}

\begin{document}
\title{Correlation-induced coherence and its use in detecting quantum phase transitions}
\author{Ming-Ming Du}
\affiliation{Department of Physics, Shandong University, Jinan 250100, China}
\author{Abdul Sattar Khan}
\affiliation{Department of Physics, Shandong University, Jinan 250100, China}
\author{Zhao-Yi Zhou}
\affiliation{Department of Physics, Shandong University, Jinan 250100, China}
\author{Da-Jian Zhang}
\email{zdj@sdu.edu.cn}
\affiliation{Department of Physics, Shandong University, Jinan 250100, China}

\date{\today}
\begin{abstract}
The past two decades have witnessed a surge of interest in exploring correlation and coherence measures to investigate quantum phase transitions (QPTs). Here, motivated by the continued push along this direction, we propose a measure which is built upon the so-called degree of coherence, and advocate using the susceptibility of our measure to detect QPTs. We show that our measure can capture both the notions of coherence and correlations exhibited in bipartite states and therefore represents a hybrid of these two notions. Through examining the XXZ model and the Kitaev honeycomb model, we demonstrate that our measure is favorable for detecting QPTs in comparison to many previous proposals.

~\\
 \textbf{correlation measures, coherence measures, quantum phase transitions}
~\\

\noindent\textbf{PACS number(s):} 03.67.-a, 64.70.Tg, 75.10.Pq
\end{abstract}

\maketitle

\section{Introduction}
Understanding quantum phase transitions (QPTs) lies at the heart of quantum many-body physics \cite{Sachdev1999}. Unlike classical phase transitions which are induced by thermal fluctuations, QPTs occur at absolute zero and are caused by quantum fluctuations originating from the Heisenberg uncertainty principle. The past two decades have witnessed a surge of interest in borrowing tools from quantum resource theories
to characterize QPTs. The best known example is the pioneering work \cite{2002Osterloh608}, in which the pairwise entanglement of two nearest neighbors, quantified by the concurrence of entanglement \cite{1998Wootters2245}, has been found to be capable of signaling the QPTs exhibited in the transverse Ising model.
This work has motivated a series of studies that make use of entanglement measures, such as entanglement of formation \cite{1998Wootters2245}, entanglement entropy \cite{1996Bennett2046}, negativity \cite{Vidal2002}, and global entanglement \cite{Meyer2002}, to study QPTs in a plethora of physical models
\cite{Osborne2002,Vidal2003,Wu2004,Refael2004,Gu2004,Anfossi2005,Gu2005,
Wei2005,Wu2006,Oliveira2006,Oliveira2006a,Buonsante2007,Sun2007,Orus2008,Facchi2008,Amico2008,
Hamma2008,Rulli2010,Pollmann2010,Chiara2012,Hofmann2014,Sahling2015,Bayat2017,Pezze2017,
Vidmar2018,Wlodzynski2020,Zhang2021,Yi2006}. Meanwhile, further effort has been devoted to establishing the links between QPTs and various other types of correlation measures \cite{Dillenschneider2008,Sarandy2009,Cui2010,Sun2010,2012Ren60305,2017Ye,2021Du12418}, such as quantum discord \cite{Ollivier2001}, Bell nonlocality \cite{Deng2012,Justino2012}, and quantum mutual information \cite{Groisman2005}.

In addition to correlations, another fundamental feature of quantum systems is coherence, which describes the capability of a quantum state to
exhibit quantum interference phenomena.
Ever since Baumgratz \textit{et al.}'s work
\cite{Baumgratz2014}, quantifying coherence has received much attention and a substantial number of coherence measures have been proposed successively \cite{Tan2016,2016Yu60302,2016Yu60303,Ma2016a,Ma2017,2018Zhang170501,2019Liu70402,2021Jin280311}. This has spurred a vibrant activity on exploring coherence measures to investigate QPTs. For example, the relative entropy of coherence has been introduced to identify QPTs in several physical models \cite{Chen2016}, the $l_1$ norm of coherence has been used to study critical properties of XY spin systems \cite{Qin2018}, and the steered quantum coherence has been proposed to signal QPTs in spin chains \cite{Hu2020,2021Hu}.
More examples involving other coherence measures, such as the total quantum coherence, the local and intrinsic quantum coherence, and the coherence measure based on the Jensen-Shannon divergence, can be found in Refs.~\cite{Radhakrishnan2017,
Mondal2017,Tan2018,Li2018,Hu2020,Xie2020,Xie2020a,Ye2020a,Ye2018c}.

While both correlation and coherence measures have been shown to be useful in detecting QPTs in a large number of systems, it has been found that each of these two kinds of measures has its own limitations for detecting QPTs. For example, the topological QPT in the Kitaev honeycomb model cannot be detected by any correlation measure adopted so far, as the local spins of this model are classically correlated \cite{Cui2010,Wang2010,Yang2008}. On the other hand, it seems difficult to use  the coherence measures in the previous works \cite{Chen2016,Qin2018,Radhakrishnan2017,
Tan2018,Li2018,Ye2020a} to detect the infinite-order Kosterlitz-Thouless QPT (KT-QPT). Specifically, the authors of Ref.~\cite{Ye2020a} have demonstrated that the correlated coherence \cite{Tan2016}  cannot reveal the KT-QPT in some critical systems; the authors of Ref.~\cite{Malvezzi2016} have found that the $l_1$ norm of coherence, the relative entropy of coherence, and the skew-information-based coherence cannot detect the KT-QPT in the spin-$1$ XXZ model; and the authors of Ref.~\cite{Ye2020} have pointed out that the relative entropy of coherence and the coherence measure based on the Jensen-Shannon divergence cannot detect the KT-QPT in the spin-$1/2$ XXZ model.

In view of the above research, we would like to shift our focus to the measures that characterize a hybrid of the two notions of correlations and coherence. We aim to ask the following question: Is there a hybrid measure that can get rid of the above limitations and provide a more effective tool for detecting QPTs? In this work, we propose a hybrid measure and advocate using its susceptibility to detect QPTs. While our measure is built upon the so-called degree of coherence \cite{Mandel1995,Patoary2019}, we show that our measure is correlation-induced since it can capture classical and quantum correlations exhibited in bipartite states. For this reason, we refer to our measure as correlation-induced coherence (CIC). Through examining the XXZ model and the Kitaev honeycomb model, we demonstrate that the CIC is favorable for detecting QPTs in comparison to many previous proposals based on correlation and coherence measures  \cite{Chen2016,Qin2018,Radhakrishnan2017,Tan2018,Li2018,Ye2020,Ye2020a}. A few attempts on using hybrid measures to detect QPTs have been carried out in Refs.~\cite{Hu2020,Ye2020a}.

This paper is organized as follows.  In Sec.~\ref{sec2}, we introduce the CIC and discuss its properties. In Sec.~\ref{sec3}, we use the CIC to detect the QPTs in the XXZ model. In Sec.~\ref{sec4}, we explore the CIC to detect the QPTs in the Kitaev honeycomb model. We conclude this work in Sec.~\ref{conclusions}.

\section{Correlation-induced coherence}\label{sec2}

Let us consider the bipartite situation that
Alice and Bob share a quantum state $\rho_{AB}$. Here, $\rho_{AB}$ is assumed to be $d\times d$-dimensional for simplicity. Our following discussion can be straightforwardly extended to the general setting that the dimension of Alice's system is different from that of Bob's system. The reduced state of Bob's system is
\begin{align}\label{rhob}
\rho_B=\tr_A\rho_{AB}.
\end{align}
Suppose that Alice performs a local measurement on her side. Note that a POVM element associated with the local measurement can be described by a positive-semidefinite operator $M$ satisfying $0\leq M\leq I$ in general, with $I$ denoting the identity operator. Here and henceforth, for two operators $A$ and $B$, we say $A\leq B$ if $B-A$ is a positive-semidefinite operator \cite{2016Zhang12117,2016Zhang52132,2019Zhang42104,2022Zhang81}.  Once the measurement outcome associated with $M$ occurs, the reduced state of Bob's system changes from the state $\rho_B$ to the state
\begin{align}\label{rhobm}
\rho_B^M={\tr_A(M\otimes I\rho_{AB})}/p_M,
\end{align}
where $p_M:=\tr(M\otimes I\rho_{AB})$ denotes the probability of getting $\rho_B^M$ \cite{2019ZWG62121,2020Zhang23418}.
Essentially, the change of the reduced state on Bob's side is due to the classical or quantum correlations exhibited in $\rho_{AB}$.
Besides, the coherence of $\rho_B^M$ is generally different from that of $\rho_B$. Combining these two facts, we have that, generally speaking, the classical or quantum correlations exhibited in $\rho_{AB}$ leads to the change of the reduced state on Bob's side, which further results in the change of the coherence of the reduced state. Therefore, we may explore the difference between the coherence of $\rho_B^M$ and that of $\rho_B$ to characterize the degree of the classical or quantum correlations exhibited in $\rho_{AB}$, which is the basic idea underpinning our proposal of the CIC.

To formalize the above idea, we need to find a way to quantify the coherence of the reduced state of Bob's system. To this end, we resort to the so-called degree of coherence \cite{Mandel1995,Patoary2019},
\begin{align}\label{DoC}
D(\rho)=\sqrt{\frac{d\tr\rho^2-1}{d-1}},
\end{align}
where $\rho$ denotes a $d$-dimensional quantum state. $D(\rho)$ has been regarded as a measure of intrinsic coherence \cite{Mandel1995,Kagalwala2012,Svozilik2015,Yao2016,Cernoch2018,Kalaga2018,
Patoary2019,Fan2019,Du2021}. Moreover, it has an operational meaning, that is, the square of $D(\rho)$ is equal to the celebrated Brukner-Zeilinger invariant information \cite{Brukner1999} up to a multiplicative factor,
\begin{align}
D(\rho)^2=\frac{d}{d-1}\mathcal{I}_{BZ},
\end{align}
where $\mathcal{I}_{BZ}$ is the Brukner-Zeilinger information \cite{Brukner1999} representing the sum of the individual measures of information over a complete set of mutually complementary observations. Using Eq.~(\ref{DoC}), we can quantify the coherence of Bob's system before and after the local measurement as $D(\rho_B)$ and $D(\rho_B^M)$, respectively. Then, the quantity $D(\rho_B^M)-D(\rho_B)$ stands for the change of the coherence of Bob's system. Note that this quantity depends on $M$ and optimizing the quantity over all possible $M$ can remove the dependence and give rise to a functional of $\rho_{AB}$. The above consideration motivates us to define the CIC as follows:
\begin{align}\label{CIC}
C^{\rightarrow}(\rho_{AB})=\max_{M} \left\{D(\rho_B^M)-D(\rho_B)\right\},
\end{align}
which, as detailed below, represents the coherence induced by the classical or quantum correlations exhibited in $\rho_{AB}$. The superscript $\rightarrow$ appearing in Eq.~(\ref{CIC}) is used to indicate that the local measurement involved is performed on Alice's side. Likewise, we can define the CIC associated with the local measurement performed on Bob's side as
\begin{align}
C^{\leftarrow}(\rho_{AB})=\max_{M} \left\{D(\rho_A^M)-D(\rho_A)\right\},
\end{align}
where $\rho_A$ and $\rho_A^M$ are defined in a way similar to $\rho_B$ and $\rho_B^M$, respectively. Note that $C^{\rightarrow}(\rho_{AB})$ and $C^{\leftarrow}(\rho_{AB})$ may or may not be identical to each other, depending on the specific state $\rho_{AB}$ under consideration. Interestingly, for $\rho_{AB}$ considered in this work,
$C^{\rightarrow}(\rho_{AB})$ is identical to $C^{\leftarrow}(\rho_{AB})$. We therefore only focus on $C^{\rightarrow}(\rho_{AB})$ in the following.

Let us move on to the discussion of the geometrical meaning of the CIC. To do this, we resort to the notion of the generalized Bloch vector \cite{2003Kimura339}. Let $\Lambda_k$, $k=1,\cdots,d^2-1$, be the generators of $SU(d)$ satisfying the conditions
\begin{align}
{\Lambda}_k={\Lambda}_k^{\dag}, ~~~\tr\Lambda_k=0,~~~  \tr({\Lambda}_j{\Lambda}_k)=2\delta_{jk}.
\end{align}
Then, we can express $\rho_{AB}$ as
\begin{align}\label{rho}
\rho_{AB}&=\frac{I}{d}\otimes\frac{I}{d}+\frac{1}{2}
\bm{a}\cdot\bm{{\Lambda}}\otimes\frac{I}{d}+\frac{1}{2} \frac{I}{d}\otimes\bm{b}\cdot\bm{{\Lambda}}\notag\\
&+\frac{1}{4}\sum_{i,j=1}^{d^{2}-1} t_{ij}{\Lambda}_{i}\otimes {\Lambda}_{j}.
\end{align}
Here, $\bm{{\Lambda}}=({\Lambda}_1,\cdot\cdot\cdot,{\Lambda}_{d^2-1})$ is the collective representation of the generators of $SU(d)$. $\bm{a}=\tr(\bm{{\Lambda}}\otimes I\rho_{AB})$ and $\bm{b}=\tr(I\otimes\bm{{\Lambda}}\rho_{AB})$ are the generalized Bloch vectors for $\rho_A$ and $\rho_B$ \cite{2003Kimura339}, respectively. $t_{ij}=\tr({\Lambda}_{i}\otimes{\Lambda}_{j}\rho_{AB})$
is responsible for the correlations exhibited in $\rho_{AB}$. Hereafter, $\bm{T}$ is used to denote the matrix with entries $t_{ij}$. Note that
\begin{align}\label{rhob-BR}
\rho_B=\frac{I}{d}+\frac{1}{2}\bm{b}\cdot\bm{{\Lambda}}.
\end{align}
Substituting Eq.~(\ref{rhob-BR}) into Eq.~(\ref{DoC}), we have
\begin{align}\label{db}
D(\rho_B)=\sqrt{\frac{d}{2(d-1)}}\abs{\bm{b}}.
\end{align}
On the other hand, $M$ can be expressed as
$M=c(I+\bm{m}\cdot\bm{{\Lambda}})$, where $c$ and $\bm{m}$ are some real parameters ensuring that $0\leq M\leq I$. Inserting $M=c(I+\bm{m}\cdot\bm{{\Lambda}})$ into Eq.~(\ref{rhobm}), we have
\begin{align}\label{rhobm2}
\rho_B^M=\frac{I}{d}+\frac{1}{2}\bm{b}_{M}\cdot\bm{{\Lambda}},
\end{align}
where
\begin{align}\label{bm}
\bm{b}_M=\frac{\bm{b}+\bm{T}^T\bm{m}}{1+\bm{a}\cdot\bm{m}}.
\end{align}
Then,
\begin{align}\label{dbm}
D(\rho_B^M)=\sqrt{\frac{d}{2(d-1)}}|\bm{b}_M|.
\end{align}
Using Eqs.~(\ref{db}) and (\ref{dbm}), we can express the CIC as
\begin{align}\label{CIC1}
C^{\rightarrow}(\rho_{AB})=\max_{\bm{m}}\sqrt{\frac{d}{2(d-1)}}
\left(\abs{\bm{b}_M}-\abs{\bm{b}}\right).
\end{align}
Note that $\abs{\bm{b}}$ and $\abs{\bm{b}_M}$ represent the lengths of the generalized Bloch vectors for $\rho_B$ and $\rho_B^M$, respectively. Equation (\ref{CIC1}) shows that $C^{\rightarrow}(\rho_{AB})$ can be interpreted as the maximal increment of the length of the generalized Bloch vector of the reduced state on Bob's side up to a multiplicative factor.

Now, we point out some interesting properties that the CIC has.

\emph{Property 1.} $C^{\rightarrow}(\rho_{AB})$  is invariant under any local unitary operation, that is, $C^{\rightarrow}(\rho_{AB})=C^{\rightarrow}(U_A\otimes U_B\rho_{AB}U_A^\dagger\otimes U_B^\dagger)$, where $U_A$ and $U_B$ are two arbitrary unitary operators acting on Alice's and Bob's system, respectively.

This property indicates that the CIC does not rely on a preferred local basis.
Notably, while some basis-independent coherence measures have been proposed \cite{2021Designolle220404,Svozilik2015,Yao2016,Cernoch2018,Kalaga2018,Fan2019,Du2021}, most of the coherence measures put forward so far are basis-dependent. The basis dependence of the coherence
measures may put the role of these
measures as detectors of QPTs into question. For example, whereas the relative entropy of coherence can detect the QPTs in the Kitaev honeycomb model if the basis is chosen to be the eigenbasis of $\sigma_z$, it cannot if the basis is chosen to be the eigenbasis of $\sigma_x$ \cite{Chen2016}.

\textit{Proof of property 1.} Under the action of the local unitary operation described by $U_A\otimes U_B$, the state $\rho_{AB}$ is transformed into the state $\tilde{\rho}_{AB}=U_A\otimes U_B\rho_{AB}U_A^\dagger\otimes U_B^\dagger$, that is,
\begin{align}\label{zhang2}
\rho_{AB}\xrightarrow{U_A\otimes U_B} \tilde{\rho}_{AB}=U_A\otimes U_B\rho_{AB}U_A^\dagger\otimes U_B^\dagger.
\end{align}
Let $\tilde{\rho}_B$ and $\tilde{\rho}_B^M$ denote the reduced states of $\tilde{\rho}_{AB}$ on Bob's side before and after performing the local measurement on Alice's side. Noting that $\tilde{\rho}_B=\tr_A\tilde{\rho}_{AB}$ and using Eq.~(\ref{zhang2}), we have
\begin{align}
\tilde{\rho}_B=U_B\rho_BU_B^\dagger.
\end{align}
Hence,
\begin{align}\label{zhang5}
D(\tilde{\rho}_B)=D(\rho_B),
\end{align}
where we have used the fact that the degree of coherence in Eq.~(\ref{DoC}) is invariant under any unitary operation. On the other hand, using the cyclic property of the trace, that is, $\tr AB=\tr BA$, for two matrices $A$ and $B$, we have
\begin{align}\label{zhang3}
\tr_A\left(M\otimes I\tilde{\rho}_{AB}\right)&=\tr_A\left(M\otimes IU_A\otimes U_B\rho_{AB}U_A^\dagger\otimes U_B^\dagger\right)\nonumber\\
&=U_B\tr_A\left(U_A^\dagger MU_A\otimes I\rho_{AB}\right) U_B^\dagger.
\end{align}
Besides,
\begin{align}\label{zhang4}
\tr\left(M\otimes I\tilde{\rho}_{AB}\right)&=\tr\left(M\otimes IU_A\otimes U_B\rho_{AB}U_A^\dagger\otimes U_B^\dagger\right)\nonumber\\
&=\tr\left(U_A^\dagger MU_A\otimes I\rho_{AB}\right).
\end{align}
Substituting Eqs.~(\ref{zhang3}) and (\ref{zhang4}) into Eq.~(\ref{rhobm}), we have
\begin{align}
\tilde{\rho}_B^M=U_B\rho_B^{U_A^\dagger MU_A}U_B^\dagger.
\end{align}
Then,
\begin{align}\label{zhang6}
D\left(\tilde{\rho}_B^M\right)=D\left(\rho_B^{U_A^\dagger MU_A}\right).
\end{align}
Inserting Eqs.~(\ref{zhang5}) and (\ref{zhang6}) into Eq.~(\ref{CIC}) yields
\begin{align}
C^\rightarrow(\tilde{\rho}_{AB})=\max_M\left\{D\left(\rho_B^{U_A^\dagger MU_A}\right)-D(\rho_B)\right\}.
\end{align}
Noting that
\begin{align}
\max_MD\left(\rho_B^{U_A^\dagger MU_A}\right)=\max_MD\left(\rho_B^{M}\right),
\end{align}
we have $C^\rightarrow(\tilde{\rho}_{AB})=C^\rightarrow({\rho}_{AB})$. This completes the proof.

\emph{Property 2.} $C^{\rightarrow}(\rho_{AB})\geq0$ for any $\rho_{AB}$ and $C^{\rightarrow}(\rho_{AB})=0$ if and only if $\rho_{AB}$ is of the form $\rho_{AB}=\rho_A\otimes\rho_B$.

This property shows that the CIC is able to capture not only quantum correlations but also classical correlations. Notably, for some physical models, e.g., the Kitaev honeycomb model, it is classical correlations rather than quantum correlations that account for QPTs. Here, by saying ``classical correlations'' and ``quantum correlations,'' we mean the classical and quantum correlations between two subsystems of these models \cite{2020Li280312}. Consequently, the QPTs in these models cannot be revealed by any measure of quantum correlations.  As demonstrated below, unlike measures of quantum correlations, the CIC may be capable of detecting the QPTs that stem from classical correlations.

\textit{Proof of property 2.} Evidently, when $M=I$, there is $\rho_B^M=\rho_B$, which implies that $D(\rho_B^M)-D(\rho_B)=0$. Using this fact and noting that $C^{\rightarrow}(\rho_{AB})\geq D(\rho_B^M)-D(\rho_B)$ for any $M$, we have that $C^{\rightarrow}(\rho_{AB})\geq0$.
It remains to show that $C^{\rightarrow}(\rho_{AB})=0$ if and only if $\rho_{AB}=\rho_A\otimes\rho_B$. The `if' part is obvious. Indeed, if $\rho_{AB}=\rho_A\otimes\rho_B$, we have $\rho_B^M=\rho_B$ for any $M$, that is, the reduced state on Bob's side does not depend on the local measurement on Alice's side. A direct consequence of this fact is that $C^{\rightarrow}(\rho_{AB})=0$. To prove the `only if' part, we deduce from Eq. (\ref{CIC1}) that we need to show
$|\bm{b}_M|\leq|\bm{b}|$ for any $M$ satisfying $0\leq M\leq I$.
Equation (\ref{bm}) allows for rewriting $|\bm{b}_M|\leq|\bm{b}|$ as
\begin{align}\label{du1}
\left|\frac{\bm{b}+\bm{T}^T\bm{m}}{1+\bm{a}\cdot\bm{m}}\right|\leq|\bm{b}|.
\end{align}
Here, $\bm{m}$ is the collective representation of $d^2-1$ real parameters satisfying $0\leq c(I+\bm{m}\cdot\bm{\Lambda})\leq I$. It follows from Eq.~(\ref{du1}) that
\begin{align}\label{du3}
|\bm{b}+\bm{T}^T\bm{m}|^2\leq(1+\bm{a}\cdot\bm{m})^2|\bm{b}|^2.
\end{align}
Using the equalities $|\bm{b}+\bm{T}^T\bm{m}|^2=(\bm{b}^T+\bm{m}^T\bm{T})(\bm{b}+\bm{T}^T\bm{m})$, $|\bm{b}|^2=\bm{b}^T\bm{b}$, and $\bm{a}\cdot\bm{m}=\bm{m}^T\bm{a}$, we can rewrite Eq.~(\ref{du3}) as
\begin{align}\label{du4}
\bm{m}^T(\bm{T}\bm{T}^T-\bm{a}\bm{a}^T\abs{\bm{b}}^2)
\bm{m}+2\bm{m}^T(\bm{T}-\bm{a}\bm{b}^T)\bm{b}\leq0.
\end{align}
Rewriting Eq.~(\ref{du4}) through replacing $\bm{m}$ by $-\bm{m}$, we have that
\begin{align}\label{du5}
\bm{m}^T(\bm{T}\bm{T}^T-\bm{a}\bm{a}^T\abs{\bm{b}}^2)\bm{m}-2\bm{m}^T(\bm{T}-\bm{a}\bm{b}^T)\bm{b}\leq0.
\end{align}
Taking the sum of the two sides of Eqs.~(\ref{du4}) and (\ref{du5}), respectively, we obtain
\begin{align}\label{du6}
\bm{m}^T(\bm{T}\bm{T}^T-\bm{a}\bm{a}^T|\bm{b}|^2)\bm{m}\leq0,
\end{align}
which is equivalent to the inequality
\begin{align}\label{du7}
\bm{T}\bm{T}^T\leq\bm{a}\bm{a}^T|\bm{b}|^2.
\end{align}
Equation (\ref{du7}) implies that the support of $\bm{T}\bm{T}^T$ must be contained in the support of $\bm{a}\bm{a}^T|\bm{b}|^2$. Here, the support of a matrix is defined to be the linear space spanned by the columns of the matrix. Besides, since the support of $\bm{a}\bm{a}^T|\bm{b}|^2$ is at most one-dimensional, the support of $\bm{T}\bm{T}^T$ must be at most one-dimensional. Therefore, we can express $\bm{T}$ as $\bm{T}=\bm{a}\bm{c}^{T}$, with $\bm{c}$ being a vector to be determined.  Substituting $\bm{T}=\bm{a}\bm{c}^{T}$ into Eq.~(\ref{du7}), we have
\begin{align}\label{du9}
|\bm{c}|\leq|\bm{b}|.
\end{align}
Moreover, substituting $\bm{T}=\bm{a}\bm{c}^T$ into Eq. (\ref{du4}), we have
\begin{align}\label{du8}
(\bm{m}^T\bm{a})^2(|\bm{c}|^2-|\bm{b}|^2)+2\bm{m}^T\bm{a}(\bm{c}^T\bm{b}-|\bm{b}|^2)\leq0.
\end{align}
The left-hand side of Eq.~(\ref{du8}) can be regarded as a quadratic equation with $\bm{m}^T\bm{a}$ representing an unknown and $(|\bm{c}|^2-|\bm{b}|^2)$ and $2(\bm{c}^T\bm{b}-|\bm{b}|^2)$ representing the coefficients. Using some basic knowledge about quadratic equations, we have that Eq.~(\ref{du8}) with Eq.~(\ref{du9}) holds if and only if
\begin{align}\label{zhang1}
\bm{c}^T\bm{b}=|\bm{b}|^2.
\end{align}
Combining Eqs.~(\ref{du9}) and (\ref{zhang1}) gives $\bm{c}=\bm{b}$. Therefore,
\begin{align}
\bm{T}=\bm{a}\bm{b}^T,
\end{align}
from which it follows that $\rho_{AB}=\rho_A\otimes\rho_B$. This completes the proof.

\emph{Property 3.} In the regime of pure states, $C^{\rightarrow}(\ket{\psi_{AB}})$ reaches the maximum value for a pure state $|\psi_{AB}\rangle$ if and only if the state $|\psi_{AB}\rangle$ is a maximally entangled pure state.

Property 3 is complementary to property 2. Indeed, while property 2 implies that $C^\rightarrow(\rho_{AB})$ attains its minimum value if there is no correlation in $\rho_{AB}$, property 3 shows that $C^\rightarrow(\rho_{AB})$ reaches its maximum value if $\rho_{AB}$ has the most entanglement.

\textit{Proof of property 3.} To prove the `if' part, we make use of the fact that  the length of the generalized Bloch vector $\bm{\lambda}$ of an arbitrary state $\rho=\frac{I}{d}+\frac{1}{2}\bm{\lambda}\cdot\bm{\Lambda}$ is confined to the interval $[0,\sqrt{2(d-1)/d}]$ \cite{2003Kimura339}, that is, $0\leq\abs{\bm{\lambda}}\leq \sqrt{2(d-1)/d}$. Hence,
$0\leq\abs{\bm{b}}\leq \sqrt{2(d-1)/d}$ and $0\leq\abs{\bm{b}_M}\leq \sqrt{2(d-1)/d}$, which further implies that
$\abs{\bm{b}_M}-\abs{\bm{b}}\leq \sqrt{2(d-1)/d}$. Using this fact and Eq.~(\ref{CIC1}), we have
\begin{align}\label{pro3-2}
C^{\rightarrow}(\rho_{AB})\leq 1,
\end{align}
for an arbitrary state $\rho_{AB}$. On the other hand, it is well-known that any maximally entangled pure state can be expressed as
\begin{align}\label{pro3-3}
|\Psi_{AB}\rangle=\frac{1}{\sqrt{d}}\sum_{i=1}^d\ket{\phi_i}_A\ket{\varphi_i}_B,
\end{align}
where $\ket{\phi_i}_A$ and $\ket{\varphi_i}_B$, $i=1,\cdots,d$, constitute two orthonormal bases for Alice's and Bob's systems, respectively. Evidently, the reduced state of $|\Psi_{AB}\rangle$ on Bob's side is the completely mixed state, i.e., $\rho_B=I/d$. This implies that
\begin{align}\label{pro3-4}
D(\rho_B)=0.
\end{align}
Besides, when $M=\ket{\phi_i}\bra{\phi_i}$, $\rho_B^M=\ket{\varphi_i}\bra{\varphi_i}$, for which there is
\begin{align}\label{pro3-5}
D(\rho_B^M)=1.
\end{align}
Inserting Eqs.~(\ref{pro3-4}) and (\ref{pro3-5}) into Eq.~(\ref{CIC}), we have that
\begin{align}\label{pro3-6}
C^\rightarrow(\ket{\Psi_{AB}})\geq 1,
\end{align}
which, in conjunction with Eq.~(\ref{pro3-2}), yields
\begin{align}\label{pro3-7}
C^\rightarrow(\ket{\Psi_{AB}})= 1.
\end{align}
Therefore, the maximum value of the CIC is $1$ and this value is reached when the state is a maximally entangled pure state. To prove the `only if' part, we need to show that the CIC cannot reach the maximum value $1$ if the bipartite pure state $|\psi_{AB}\rangle$ under consideration is not a maximally entangled pure state. To this end, we resort to the fact that any bipartite pure state $|\psi_{AB}\rangle$ can be written as
\begin{align}
|\psi_{AB}\rangle=\sum_{i=1}^d \sqrt{p_i}\ket{\phi_i}_A\ket{\varphi_i}_B,
\end{align}
where $p_i$ are non-negative real numbers satisfying $\sum_{i=1}^d p_i=1$ known as Schmidt
coefficients. Apparently,
\begin{align}\label{pro3-8}
\rho_B=\sum_{i=1}^{d}p_i\ket{\varphi_i}\bra{\varphi_i}.
\end{align}
Then, there is $\tr\rho_B^2=\sum_{i=1}^{d}p_i^2$. It is easy to see that $\tr\rho_B^2> 1/d$, since $|\psi_{AB}\rangle$  is not a maximally entangled pure state. Noting that $\tr\rho_B^2=\frac{1}{d}+\frac{1}{2}\abs{\bm{b}}^2$, we have  $\abs{\bm{b}}> 0$, which further leads to
\begin{align}\label{pro3-9}
\abs{\bm{b}_M}-\abs{\bm{b}}<\sqrt{\frac{2(d-1)}{d}},
\end{align}
by noting that $\abs{\bm{b}_M}\leq \sqrt{2(d-1)/d}$ in general. Equation (\ref{pro3-9}) shows that
\begin{align}
C^\rightarrow(|\psi_{AB}\rangle)<1,
\end{align}
for any $|\psi_{AB}\rangle$ that is not maximally entangled. This completes the proof.

\section{Detecting the QPTs in the XXZ model}\label{sec3}

Here we examine the XXZ model, which is a spin-$1/2$ chain consisting of spins interacting with each other through anisotropic Heisenberg interactions. Its Hamiltonian reads
\begin{align}\label{H-model}
H_\textrm{XXZ}=\sum_{j=1}^{N}S_{j}^{x} S_{j+1}^{x}+S_{j}^{y} S_{j+1}^{y}+\Delta S_{j}^{z} S_{j+1}^{z},
\end{align}
where $N$ denotes the number of sites, $S_{j}^{\alpha}=\sigma_{j}^\alpha/2$, $\alpha=x, y, z$, are the Pauli spin-$1/2$ operators acting on site $j$, and $\Delta$ is the so-called anisotropy parameter. This model has three phases:

(i) When $\Delta <-1$, the system is in the ferromagnetic phase, in which all the spins point in a same direction.

(ii) When $-1<\Delta<1$, the system is in the gapless phase, in which the correlations among the spins decay polynomially.

(iii) When $\Delta>1$, the system is in the antiferromagnetic phase.

The
QPTs occur at the boundaries between different phases. Specifically, the first-order QPT occurs at the critical point $\Delta=-1$, and the KT-QPT occurs at the critical point $\Delta=1$.

In what follows, we explore the CIC to detect the QPTs. To do this,
we resort to the CIC associated with two nearest-neighbor spins in this model, say, $i$ and $i+1$. That is, we are interested in  $C^\rightarrow(\rho_{i,i+1})$ for the density operator $\rho_{i,i+1}$ of spins $i$ and $i+1$. The reason of doing so is as follows. On one hand, it is known that the
QPTs exhibited in a many-body system are typically associated with the nonanalyticities in
the ground-state energy of the system \cite{Sachdev1999}. On the other hand, it has been shown that for a many-body system that involves only nearest-neighbor interactions, the ground-state energy of the system can be obtained from the reduced density matrix of two nearest-neighbor constituents \cite{Wu2004}; that is, the ground-state energy is a function of the reduced density operator $\rho_{i,i+1}$. Combining these two facts and noting that the XXZ model (as well as the Kitaev honeycomb model to be discussed later on) only involves nearest-neighbor interactions, we anticipate that QPTs are generally connected to the nonanalyticities of $\rho_{i,i+1}$ and hence the nonanalyticities of $C^\rightarrow(\rho_{i,i+1})$. Such a connection allows us to reveal QPTs through identifying the nonanalyticities of $C^\rightarrow(\rho_{i,i+1})$.

To figure out $C^\rightarrow(\rho_{i,i+1})$, we need to know the expression of $\rho_{i,i+1}$ as described by Eq.~(\ref{rho}).
It has been shown \cite{Justino2012,2021Du12418} that owing to some symmetries of Hamiltonian (\ref{H-model}), there are
\begin{align}\label{sc1}
\langle\sigma_{i}^{x}\rangle=\langle\sigma_{i}^{y}\rangle=
\langle\sigma_{i}^{z}\rangle=0,
\end{align}
\begin{align}\label{sc2}
\langle\sigma_{i}^{x} \sigma_{j}^{y}\rangle=\langle\sigma_{i}^{x} \sigma_{j}^{z}\rangle=\langle\sigma_{i}^{y} \sigma_{j}^{z}\rangle=0,
\end{align}
\begin{align}\label{sc3}
\expval{\sigma_i^x\sigma_{i+1}^x}=\expval{\sigma_i^y\sigma_{i+1}^y},
\end{align}
where the expectation values are taken in the ground state of the Hamiltonian. Taking into account symmetry constraints (\ref{sc1}), (\ref{sc2}), and (\ref{sc3}) and noting that $\langle A\rangle=\tr(A\rho_{i,i+1})$ for an operator $A$ only acting on spins $i$ and $i+1$, we have
\begin{align}\label{XXZ-ab}
\bm{a}=(0,0,0),~~~\bm{b}=(0,0,0),
\end{align}
and
\begin{align}\label{XXZ-T}
\bm{T}=\begin{pmatrix}
    \expval{\sigma_i^x\sigma_{i+1}^x} & 0 & 0 \\
    0 & \expval{\sigma_i^x\sigma_{i+1}^x} & 0 \\
    0 & 0 & \expval{\sigma_i^z\sigma_{i+1}^z}
  \end{pmatrix}.
\end{align}
Substituting Eqs.~(\ref{XXZ-ab}) and (\ref{XXZ-T}) into Eq.~(\ref{CIC1}), we obtain
\begin{align}\label{cic-XXZ}
&&C^\rightarrow(\rho_{i,i+1})
=\max_{\theta}\sqrt{\langle\sigma_i^x\sigma_{i+1}^x\rangle^2
\sin^2{\theta}+\langle\sigma_i^z\sigma_{i+1}^z\rangle^2\cos^2{\theta}}.
\end{align}
Here, we have used the fact that $\bm{m}=(\sin\theta\cos\varphi,\sin\theta\sin\varphi,\cos\theta)$ is a unit vector in $\mathbb{R}^3$, with $\theta$ and $\varphi$ denoting the polar and azimuth angles, respectively. Evidently, when $|\langle\sigma_i^x\sigma_{i+1}^x\rangle|>|\langle\sigma_i^z\sigma_{i+1}^z\rangle|$, $C^\rightarrow(\rho_{i,i+1})=|\langle\sigma_i^x\sigma_{i+1}^x\rangle|$ and the optimal
$\theta$ is given by $\theta=\pi/2$, and  when $|\langle\sigma_i^x\sigma_{i+1}^x\rangle|<|\langle\sigma_i^z\sigma_{i+1}^z\rangle|$, $C^\rightarrow(\rho_{i,i+1})=|\langle\sigma_i^z\sigma_{i+1}^z\rangle|$ and the optimal
$\theta$ is given by $\theta=0$ or $\pi$.

\begin{figure}
\includegraphics[width=0.45\textwidth]{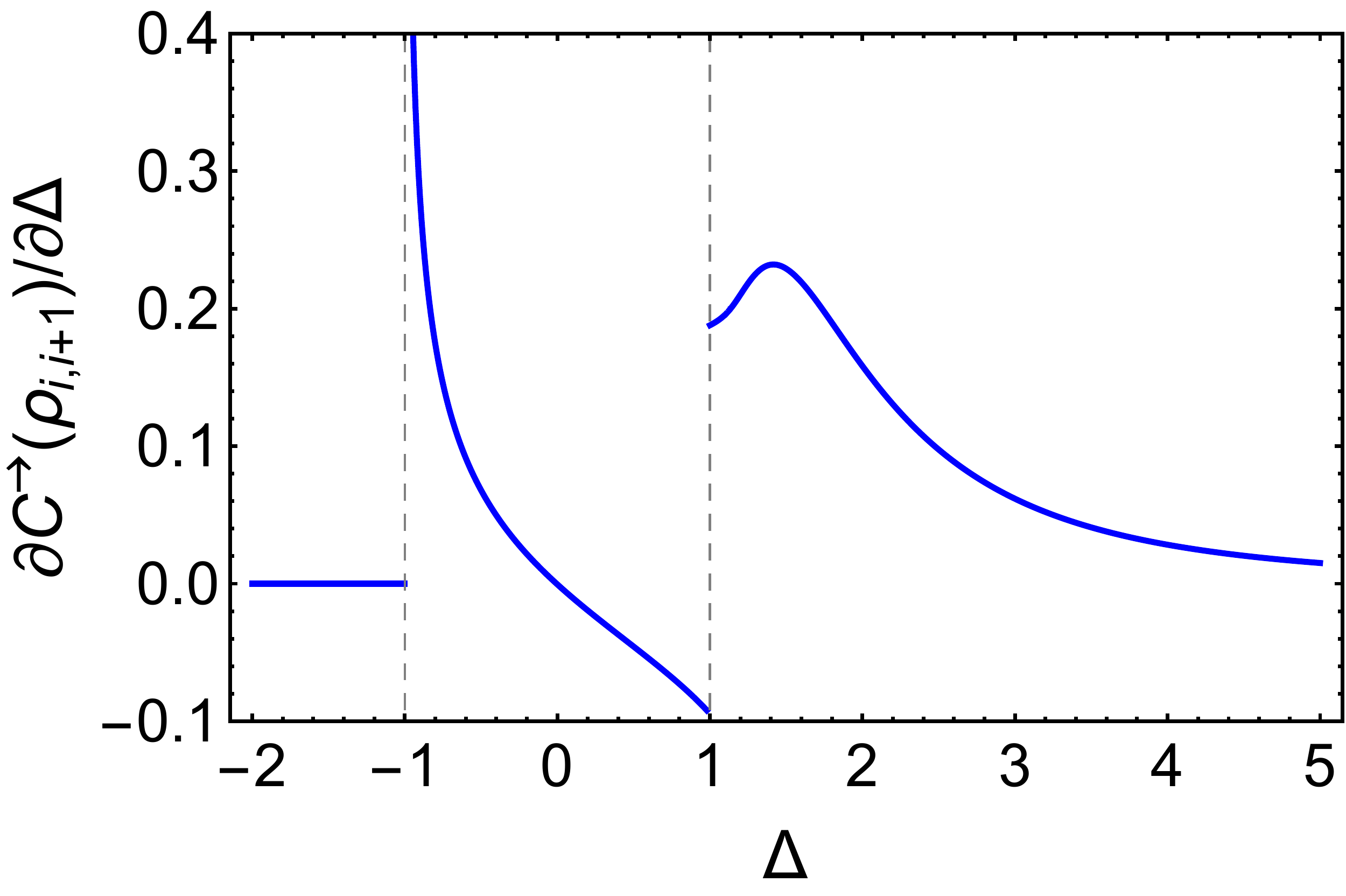}
\caption{Susceptibility $\partial C^\rightarrow(\rho_{i,i+1})/\partial\Delta$ of the correlation-induced coherence for two nearest-neighbor spins in the XXZ model.}
\label{fig1}
\end{figure}

To reveal the nonanalyticities of $C^\rightarrow(\rho_{i,i+1})$, we make use of the susceptibility of $C^\rightarrow(\rho_{i,i+1})$ defined as $\partial C^\rightarrow(\rho_{i,i+1})/\partial \Delta$. Note that it has been shown in Ref.~\cite{Shiroishi2005} that $\langle\sigma_{i}^{\alpha} \sigma_{i+1}^{\alpha}\rangle$, $\alpha=x,y,z$, can be expressed as
\begin{align}
&\langle\sigma_{i}^{x} \sigma_{i+1}^{x}\rangle=\expval{\sigma_i^y\sigma_{i+1}^y}=\frac{1}{2}\left[4 e_{g}(\Delta)-\Delta\langle\sigma_{i}^{z} \sigma_{i+1}^{z}\rangle\right],\label{cc1}\\
&\langle\sigma_{i}^{z} \sigma_{i+1}^{z}\rangle=4 \frac{\partial e_{g}(\Delta)}{\partial \Delta},\label{cc2}
\end{align}
where $e_g(\Delta)$ denotes the ground state energy, given by \cite{Yang1966,Yang1966a}
\begin{align}\label{energy}
e_{g}(\Delta)=\left\{\begin{array}{lc}
-\frac{\Delta}{4}, & \Delta \leqslant-1, \\
\frac{\Delta}{4}+\frac{\sin \pi \xi}{2 \pi} \int_{-\infty+\frac{i}{2}}^{\infty+\frac{i}{2}} d x \frac{1}{\sinh x} \frac{\cosh \xi x}{\sinh \xi x}, & -1<\Delta<1, \\
\frac{1}{4}-\ln 2, & \Delta=1,
\end{array}\right.
\end{align}
with $\Delta=\cos \pi \xi$. When $\Delta>1$, $e_g(\Delta)$ is obtained by changing $\xi=i \phi$ in Eq.~(\ref{energy}). With the aid of Eqs.~(\ref{cc1}) and (\ref{cc2}), we are able to numerically compute $\partial C^\rightarrow(\rho_{i,i+1})/\partial \Delta$ for different $\Delta$. The numerical results are shown in Fig.~\ref{fig1}. As can be seen from this figure, $\partial C^\rightarrow(\rho_{i,i+1})/\partial \Delta$ abruptly changes its value at $\Delta=\pm 1$, indicating that $C^\rightarrow(\rho_{i,i+1})$ displays nonanalyticities at the two points $\Delta=-1$ and $\Delta=1$. Noting that these two points correspond to the first-order QPT and the KT-QPT, respectively, we deduce that the CIC is able to detect the QPTs.

We remark that the optimization process appearing in the definition of the CIC plays an important role in detecting the KT-QPT. Indeed, from Eq.~(\ref{cic-XXZ}), we see that, in the gapless phase for which $|\sigma_i^x\sigma_{i+1}^x|>|\sigma_i^z\sigma_{i+1}^z|$,  there is $C^\rightarrow(\rho_{AB})=|\sigma_i^x\sigma_{i+1}^x|$, and in the antiferromagnetic phase for which $|\sigma_i^x\sigma_{i+1}^x|<|\sigma_i^z\sigma_{i+1}^z|$, there is $C^\rightarrow(\rho_{AB})=|\sigma_i^z\sigma_{i+1}^z|$. Hence, the nonanalyticity of $C^\rightarrow(\rho_{i,i+1})$ at $\Delta=1$ stems from the maximization process in Eq.~(\ref{cic-XXZ}). Notably, it is expected that taking optimization usually leads to some negative results in the ability of a measure to signal QPTs \cite{Justino2012}, since accidental nonanalytical behaviors may appear or disappear due to the optimization process involved in the measure. An example demonstrating this point is given in Ref.~\cite{2005Yang30302}, where it has been found that there exists a discontinuity in the susceptibility of the concurrence at which there is, however, no QPT. Interestingly, contrary to the expectation \cite{Justino2012}, the optimization process involved in the definition of the CIC results in the success of detecting the KT-QPT.

\section{Detecting the QPTs in the Kitaev honeycomb model} \label{sec4}

\begin{figure}
\includegraphics[width=0.45\textwidth]{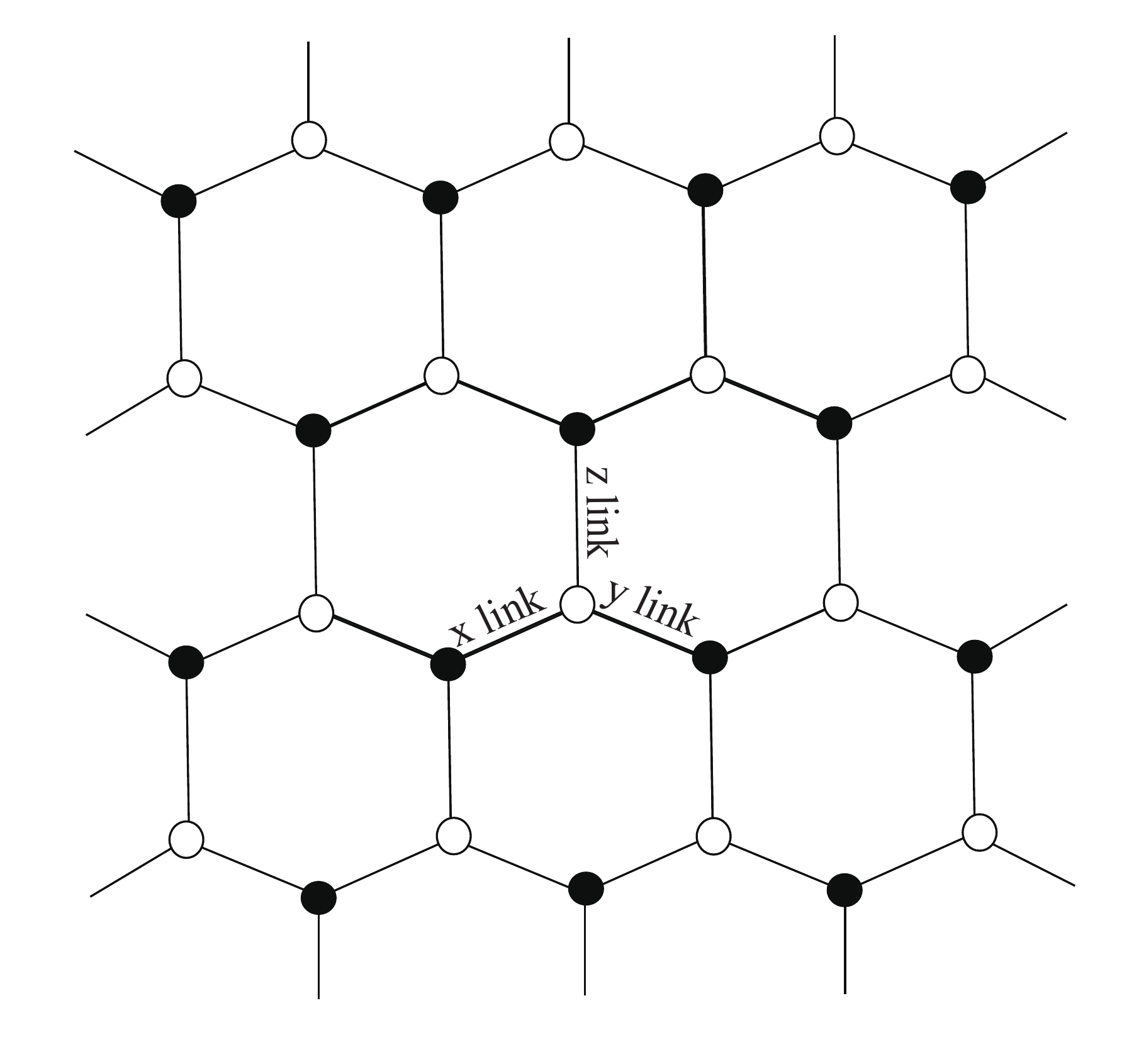}
\caption{Sketch of the Kitaev honeycomb model. Spins of $1/2$ are located at the vertices of a honeycomb lattice, which is composed of two simple sublattices denoted by empty circles and dark dots, respectively. There are three types of bonds between two of the spins labeled as x links, y links, and z links.}
\label{fig2}
\end{figure}

Here we examine the Kitaev honeycomb model. This model is two-dimensional and comprised of spins of 1/2 which are located at the vertices of a honeycomb lattice (see Fig.~\ref{fig2}). Each spin interacts with three
nearest neighbors through three distinct
bonds labeled as an $x$ link, a $y$ link, and a $z$ link. The interaction has a different coupling constant for a different bond.
The Hamiltonian of the model is \cite{Kitaev2006}
\begin{align}\label{KH-model}
H =-J_{x}\sum_{x~ \text{links}}\sigma_{j}^{x}\sigma_{k}^{x}-J_{y}\sum_{y~ \text{links}} \sigma_{j}^{y}\sigma_{k}^{y}-J_{z}\sum_{z~ \text{links}}\sigma_{j}^{z}\sigma_{k}^{z},
\end{align}
where $J_{\alpha}$, $\alpha=x, y, z$, are dimensionless coupling constants, the subindexes $j$, $k$ denote the locations of sites, and $\sigma_j^{\alpha}$ is the Pauli matrix acting at site $j$. This model has two phases \cite{Kitaev2006}:

(a) In the region of $|J_x|\leq|J_y|+|J_z|$, $|J_y|\leq|J_x|+|J_z|$, and $|J_z|\leq|J_x|+|J_y|$, it is gapless with non-Abelian excitation.

(b) In other regions, it is gapped with Abelian anyon excitations.

\begin{figure}
\includegraphics[width=0.45\textwidth]{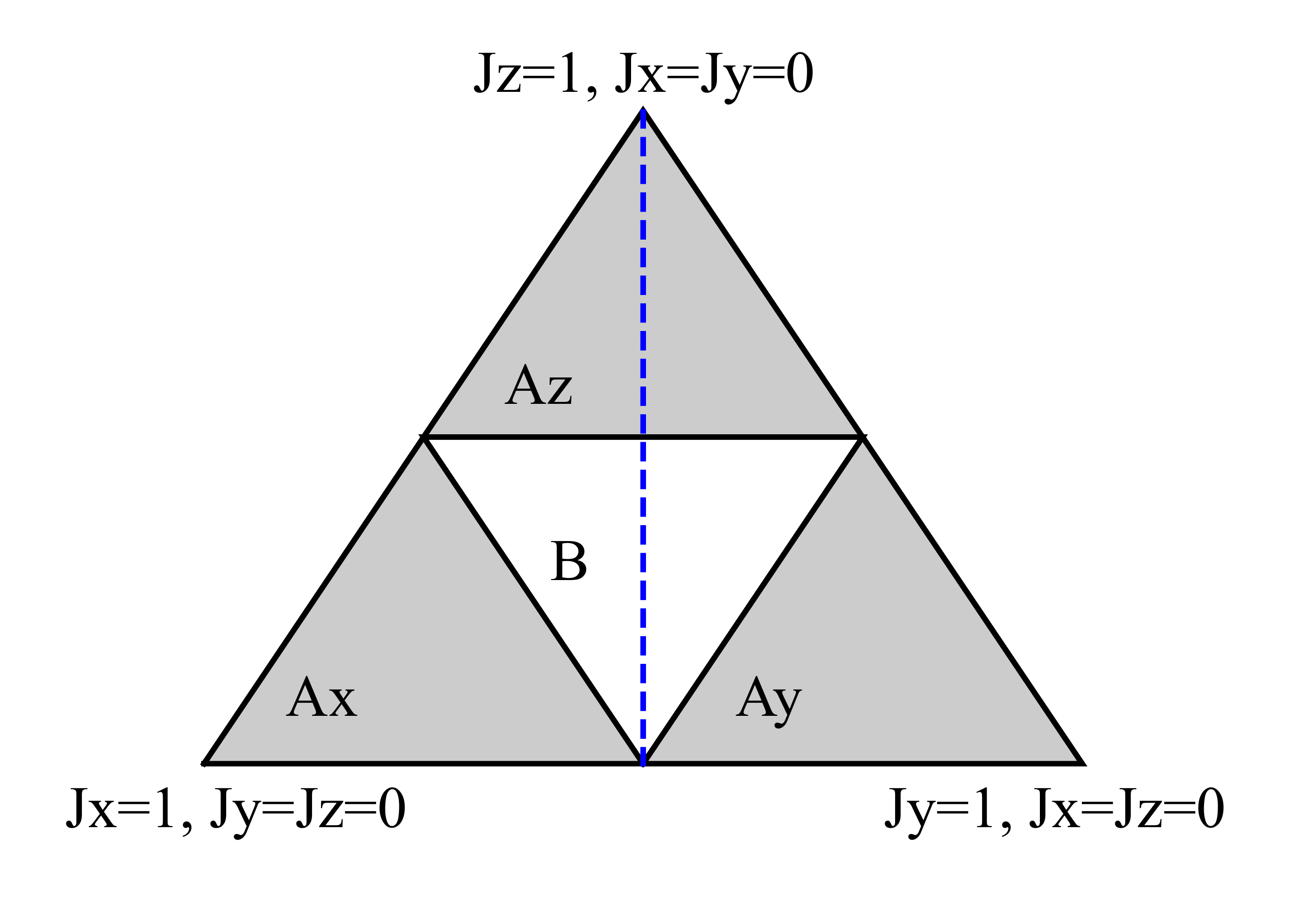}
\caption{Phase diagram of the Kitaev honeycomb model. The triangle is the section of the positive octant by the plane $J_{x}+J_{y}+J_{z}=1$. In the unshaded area labeled by $B$, the system is gapless with non-Abelian excitation, and in the three shaded areas labeled by $A_{x}, A_{y}$, and $A_{z}$, the system is gapped with Abelian anyon excitation. The blue dashed line is parameterized by  $J_{x}=J_{y}=(1-J_{z})/2$, where the topological quantum phase transition occurs at the critical point specified by $J_{z}=1/2$. }
\label{fig3}
\end{figure}

Throughout the rest of this section, following Kitaev's seminal work \cite{Kitaev2006}, we focus on the triangle in the plane $J_{x}+J_{y}+J_{z}=1$ which is schematically shown in Fig.~\ref{fig3}. The gapless and gapped phases correspond to the unshaded and shaded areas, respectively.
The topological QPTs occur whenever boundaries between different areas are crossed.

Now, let us figure out the CIC for two nearest spins in the model. Associated to each link, there is a two-site reduced state $\rho_{i,i+1}$, where $i$ and $i+1$ denote the two ends of the link in question. Note that the expression of $\rho_{i,i+1}$ is different if the type of the link is different. It has been shown that all the two-site correlation functions are zero except for the one along the direction of the link under consideration \cite{Cui2010}. Accordingly, taking the $z$ link as an example, we can express the corresponding $\rho_{i,i+1}$ as \cite{Wang2010}
\begin{align}\label{rho-KM-z}
\rho_{i,i+1}=
\frac{1}{4}\left(I\otimes I+\langle\sigma_{i}^z\sigma_{i+1}^z\rangle\sigma_{i}^z\sigma_{i+1}^z\right).
\end{align}
It then follows that
\begin{align}\label{K-ab}
\bm{a}=(0,0,0),~~~\bm{b}=(0,0,0),
\end{align}
and
\begin{align}\label{K-T}
\bm{T}=\begin{pmatrix}
    0 & 0 & 0 \\
    0 & 0 & 0 \\
    0 & 0 & \expval{\sigma_i^z\sigma_{i+1}^z}
  \end{pmatrix}.
\end{align}
Pluging Eqs.~(\ref{K-ab}) and (\ref{K-T}) into Eq. (\ref{CIC1}), we have that
\begin{align}\label{CIC-z-link}
C^{\rightarrow}
(\rho_{i,i+1})=|\langle\sigma_{i}^z\sigma_{i+1}^z\rangle|,
\end{align}
for the $z$ link. Note that it has been found \cite{Cui2010,Yang2008,Wang2010} that
\begin{align}\label{ming3}
\langle\sigma_{i}^z\sigma_{i+1}^z\rangle=\frac{1}{4 \pi^{2}} \int_{-\pi}^{\pi} \int_{-\pi}^{\pi} \frac{\epsilon}{\sqrt{\epsilon^{2}+\delta^{2}}} d \omega_{x} d \omega_{y}
\end{align}
in the thermodynamic limit,
where
$\epsilon=J_{z}+J_{x} \cos\omega_{x}+J_{y}\cos \omega_{y}$
and $\delta=J_{x} \sin \omega_{x}+J_{y} \sin \omega_y$. The explicit expression of $C^{\rightarrow}
(\rho_{i,i+1})$ can be obtained by inserting Eq.~(\ref{ming3}) into Eq.~(\ref{CIC-z-link}).

\begin{figure}
\includegraphics[width=0.45\textwidth]{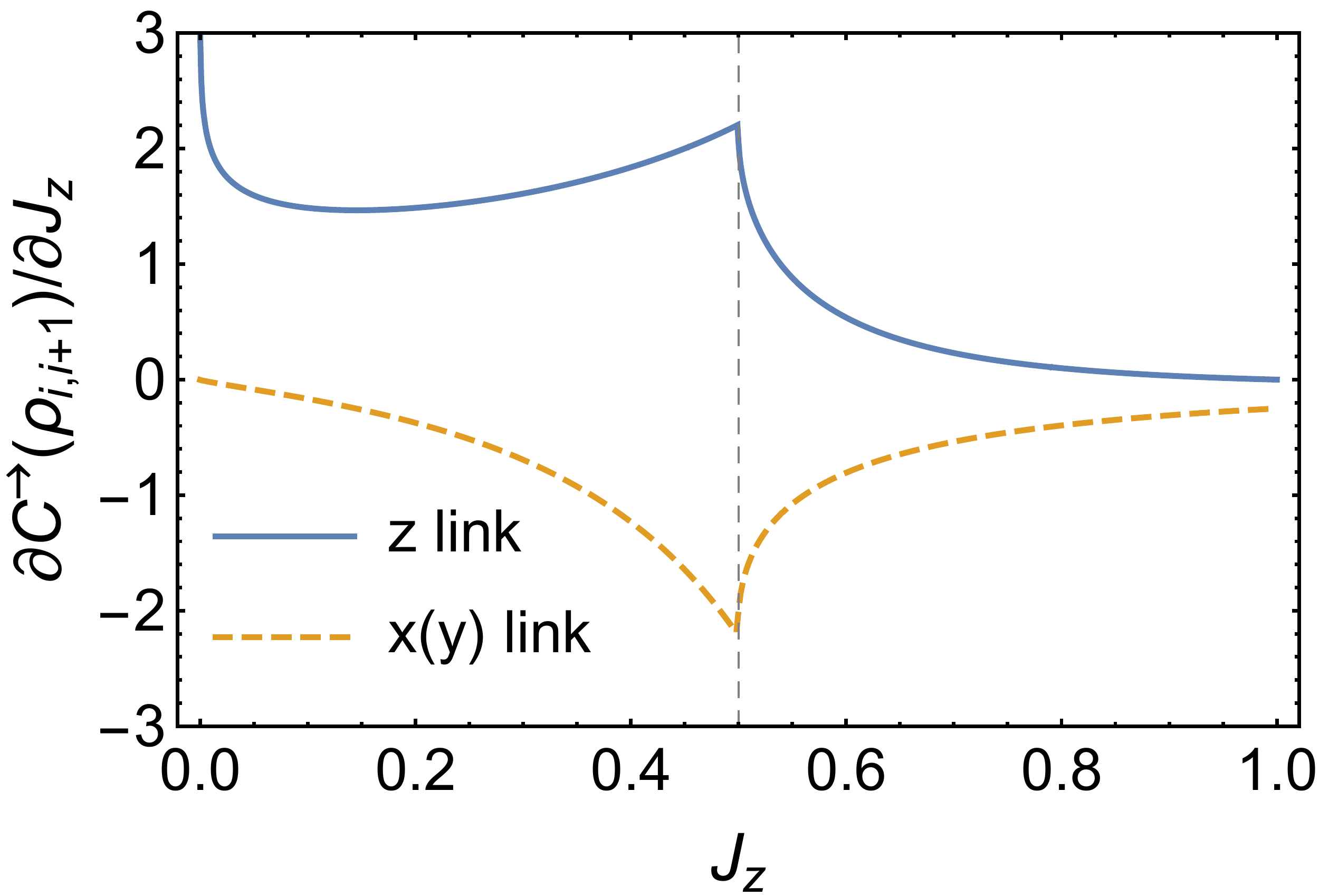}
\caption{Susceptibility $\partial C^{\rightarrow}(\rho_{i,i+1})/\partial J_z$ of the correlation-induced coherence for the three types of links in the Kitaev honeycomb model. Here,  the parameters involved satisfy $J_x=J_y=(1-J_z)/2$, corresponding to the blue dashed line in Fig.~\ref{fig3}.}
\label{fig4}
\end{figure}

To demonstrate the usefulness of the CIC in detecting the topological QPTs, we investigate
the phase transition from the gapless
phase to a gapped phase along the blue dashed line in Fig.~\ref{fig3}. This line is parameterized by  $J_{x}=J_{y}=(1-J_{z})/2$, for which the topological QPT occurs at $J_{z}=1/2$. Aided by Eq.~(\ref{CIC-z-link}) with Eq.~(\ref{ming3}), we numerically compute the susceptibility $\partial C^{\rightarrow}(\rho_{i,i+1})/\partial J_z$ of the CIC for the $z$ link (see the solid curve in Fig.~\ref{fig4}). It can be seen from Fig.~\ref{fig4} that the susceptibility of the CIC displays nonanalytical behavior at $J_z=1/2$, which is exactly the critical point where the topological QPT occurs. This clearly demonstrates that the CIC is able to detect the topological QPT. It is interesting to observe that there are only classical correlations for the state $\rho_{i,i+1}$ in Eq.~(\ref{rho-KM-z}), as this state is diagonalizable in a local basis. Therefore, any measure of quantum correlations for the state is zero and unable to reveal the topological QPTs. The reason for the success of the CIC in detecting the topological QPTs is that the CIC can capture not only quantum correlations but also classical correlations. We remark that the CIC is distinct from the so-called quantum correlated coherence \cite{Tan2016}, which only measures the quantum part of correlations and therefore is unable to detect the QPTs of the Kitaev honeycomb model \cite{2018Zhang52101}.

Besides, we have numerically computed the susceptibility $\partial C^{\rightarrow}(\rho_{i,i+1})/\partial J_z$ of the CIC for the x and y links (see the dashed curve in Fig.~\ref{fig4}).
The numerical results are consistent with that for the $z$ link, indicating that the ability of the CIC in detecting QPTs is irrespective of the specific choice of a link. Moreover, apart from investigating the phase transition along the blue dashed line in Fig.~\ref{fig3}, we have extensively examined the phase transitions along other lines. The numerical results (not shown here) suggest that the CIC is universally effective in detecting the topological QPTs in the Kitaev honeycomb model.

\section{Conclusions}\label{conclusions}
While correlation and coherence measures are useful in detecting QPTs, it has been found that each of these two kinds of measures has its own limitations for detecting QPTs. We have made an attempt to get rid of the limitations by introducing a measure that characterizes a hybrid of the two notions, namely, correlations and coherence. The introduced measure, called the CIC throughout this work, is built upon the degree of coherence but is induced by the classical and quantum correlations exhibited in bipartite states. We have demonstrated that the nonanalyticities of the CIC, which are manifested in the singular behaviors of the susceptibility, can serve as an effective and reliable detector of QPTs. We have used the detector to successfully reveal the KT-QPT in the XXZ model and the topological QPTs in the Kitaev honeycomb model. The former QPT has been found in previous works to be challenging to be revealed by most of the coherence measures proposed so far, and the latter QPT cannot be detected by any measure of quantum correlations. The present study may be seen as an example clearly demonstrating the usefulness of hybrid measures in detecting QPTs, which, we hope, can stimulate further effort.

\begin{acknowledgments}
We acknowledge support from the National Natural Science Foundation of China through Grant Nos. 11775129 and 12174224.
\end{acknowledgments}


\begin{thebibliography}{95}%
	\makeatletter
	\providecommand \@ifxundefined [1]{%
		\@ifx{#1\undefined}
	}%
	\providecommand \@ifnum [1]{%
		\ifnum #1\expandafter \@firstoftwo
		\else \expandafter \@secondoftwo
		\fi
	}%
	\providecommand \@ifx [1]{%
		\ifx #1\expandafter \@firstoftwo
		\else \expandafter \@secondoftwo
		\fi
	}%
	\providecommand \natexlab [1]{#1}%
	\providecommand \enquote  [1]{``#1''}%
	\providecommand \bibnamefont  [1]{#1}%
	\providecommand \bibfnamefont [1]{#1}%
	\providecommand \citenamefont [1]{#1}%
	\providecommand \href@noop [0]{\@secondoftwo}%
	\providecommand \href [0]{\begingroup \@sanitize@url \@href}%
	\providecommand \@href[1]{\@@startlink{#1}\@@href}%
	\providecommand \@@href[1]{\endgroup#1\@@endlink}%
	\providecommand \@sanitize@url [0]{\catcode `\\12\catcode `\$12\catcode
		`\&12\catcode `\#12\catcode `\^12\catcode `\_12\catcode `\%12\relax}%
	\providecommand \@@startlink[1]{}%
	\providecommand \@@endlink[0]{}%
	\providecommand \url  [0]{\begingroup\@sanitize@url \@url }%
	\providecommand \@url [1]{\endgroup\@href {#1}{\urlprefix }}%
	\providecommand \urlprefix  [0]{URL }%
	\providecommand \Eprint [0]{\href }%
	\providecommand \doibase [0]{http://dx.doi.org/}%
	\providecommand \selectlanguage [0]{\@gobble}%
	\providecommand \bibinfo  [0]{\@secondoftwo}%
	\providecommand \bibfield  [0]{\@secondoftwo}%
	\providecommand \translation [1]{[#1]}%
	\providecommand \BibitemOpen [0]{}%
	\providecommand \bibitemStop [0]{}%
	\providecommand \bibitemNoStop [0]{.\EOS\space}%
	\providecommand \EOS [0]{\spacefactor3000\relax}%
	\providecommand \BibitemShut  [1]{\csname bibitem#1\endcsname}%
	\let\auto@bib@innerbib\@empty
	\bibitem [{\citenamefont {Sachdev}(1999)}]{Sachdev1999}%
	\BibitemOpen
	\bibfield  {author} {\bibinfo {author} {\bibfnamefont {S.}~\bibnamefont
			{Sachdev}},\ }\href@noop {} {\emph {\bibinfo {title} {Quantum Phase
				Transitions}}}\ (\bibinfo  {publisher} {Cambridge University Press},\
	\bibinfo {address} {Cambridge New York},\ \bibinfo {year} {1999})\BibitemShut
	{NoStop}%
	\bibitem [{\citenamefont {Osterloh}\ \emph {et~al.}(2002)\citenamefont
		{Osterloh}, \citenamefont {Amico}, \citenamefont {Falci},\ and\ \citenamefont
		{Fazio}}]{2002Osterloh608}%
	\BibitemOpen
	\bibfield  {author} {\bibinfo {author} {\bibfnamefont {A.}~\bibnamefont
			{Osterloh}}, \bibinfo {author} {\bibfnamefont {L.}~\bibnamefont {Amico}},
		\bibinfo {author} {\bibfnamefont {G.}~\bibnamefont {Falci}}, \ and\ \bibinfo
		{author} {\bibfnamefont {R.}~\bibnamefont {Fazio}},\ }\href {\doibase
		10.1038/416608a} {\bibfield  {journal} {\bibinfo  {journal} {Nature
				(London)}\ }\textbf {\bibinfo {volume} {416}},\ \bibinfo {pages} {608}
		(\bibinfo {year} {2002})}\BibitemShut {NoStop}%
	\bibitem [{\citenamefont {Wootters}(1998)}]{1998Wootters2245}%
	\BibitemOpen
	\bibfield  {author} {\bibinfo {author} {\bibfnamefont {W.~K.}\ \bibnamefont
			{Wootters}},\ }\href {\doibase 10.1103/PhysRevLett.80.2245} {\bibfield
		{journal} {\bibinfo  {journal} {Phys. Rev. Lett.}\ }\textbf {\bibinfo
			{volume} {80}},\ \bibinfo {pages} {2245} (\bibinfo {year}
		{1998})}\BibitemShut {NoStop}%
	\bibitem [{\citenamefont {Bennett}\ \emph {et~al.}(1996)\citenamefont
		{Bennett}, \citenamefont {Bernstein}, \citenamefont {Popescu},\ and\
		\citenamefont {Schumacher}}]{1996Bennett2046}%
	\BibitemOpen
	\bibfield  {author} {\bibinfo {author} {\bibfnamefont {C.~H.}\ \bibnamefont
			{Bennett}}, \bibinfo {author} {\bibfnamefont {H.~J.}\ \bibnamefont
			{Bernstein}}, \bibinfo {author} {\bibfnamefont {S.}~\bibnamefont {Popescu}},
		\ and\ \bibinfo {author} {\bibfnamefont {B.}~\bibnamefont {Schumacher}},\
	}\href {\doibase 10.1103/PhysRevA.53.2046} {\bibfield  {journal} {\bibinfo
			{journal} {Phys. Rev. A}\ }\textbf {\bibinfo {volume} {53}},\ \bibinfo
		{pages} {2046} (\bibinfo {year} {1996})}\BibitemShut {NoStop}%
	\bibitem [{\citenamefont {Vidal}\ and\ \citenamefont
		{Werner}(2002)}]{Vidal2002}%
	\BibitemOpen
	\bibfield  {author} {\bibinfo {author} {\bibfnamefont {G.}~\bibnamefont
			{Vidal}}\ and\ \bibinfo {author} {\bibfnamefont {R.~F.}\ \bibnamefont
			{Werner}},\ }\href {\doibase 10.1103/PhysRevA.65.032314} {\bibfield
		{journal} {\bibinfo  {journal} {Phys. Rev. A}\ }\textbf {\bibinfo {volume}
			{65}},\ \bibinfo {pages} {032314} (\bibinfo {year} {2002})}\BibitemShut
	{NoStop}%
	\bibitem [{\citenamefont {Meyer}\ and\ \citenamefont
		{Wallach}(2002)}]{Meyer2002}%
	\BibitemOpen
	\bibfield  {author} {\bibinfo {author} {\bibfnamefont {D.~A.}\ \bibnamefont
			{Meyer}}\ and\ \bibinfo {author} {\bibfnamefont {N.~R.}\ \bibnamefont
			{Wallach}},\ }\href {\doibase 10.1063/1.1497700} {\bibfield  {journal}
		{\bibinfo  {journal} {J. Math. Phys.}\ }\textbf {\bibinfo {volume} {43}},\
		\bibinfo {pages} {4273} (\bibinfo {year} {2002})}\BibitemShut {NoStop}%
	\bibitem [{\citenamefont {Osborne}\ and\ \citenamefont
		{Nielsen}(2002)}]{Osborne2002}%
	\BibitemOpen
	\bibfield  {author} {\bibinfo {author} {\bibfnamefont {T.~J.}\ \bibnamefont
			{Osborne}}\ and\ \bibinfo {author} {\bibfnamefont {M.~A.}\ \bibnamefont
			{Nielsen}},\ }\href {\doibase 10.1103/physreva.66.032110} {\bibfield
		{journal} {\bibinfo  {journal} {Phys. Rev. A}\ }\textbf {\bibinfo {volume}
			{66}},\ \bibinfo {pages} {032110} (\bibinfo {year} {2002})}\BibitemShut
	{NoStop}%
	\bibitem [{\citenamefont {Vidal}\ \emph {et~al.}(2003)\citenamefont {Vidal},
		\citenamefont {Latorre}, \citenamefont {Rico},\ and\ \citenamefont
		{Kitaev}}]{Vidal2003}%
	\BibitemOpen
	\bibfield  {author} {\bibinfo {author} {\bibfnamefont {G.}~\bibnamefont
			{Vidal}}, \bibinfo {author} {\bibfnamefont {J.~I.}\ \bibnamefont {Latorre}},
		\bibinfo {author} {\bibfnamefont {E.}~\bibnamefont {Rico}}, \ and\ \bibinfo
		{author} {\bibfnamefont {A.}~\bibnamefont {Kitaev}},\ }\href {\doibase
		10.1103/physrevlett.90.227902} {\bibfield  {journal} {\bibinfo  {journal}
			{Phys. Rev. Lett.}\ }\textbf {\bibinfo {volume} {90}},\ \bibinfo {pages}
		{227902} (\bibinfo {year} {2003})}\BibitemShut {NoStop}%
	\bibitem [{\citenamefont {Wu}\ \emph {et~al.}(2004)\citenamefont {Wu},
		\citenamefont {Sarandy},\ and\ \citenamefont {Lidar}}]{Wu2004}%
	\BibitemOpen
	\bibfield  {author} {\bibinfo {author} {\bibfnamefont {L.-A.}\ \bibnamefont
			{Wu}}, \bibinfo {author} {\bibfnamefont {M.~S.}\ \bibnamefont {Sarandy}}, \
		and\ \bibinfo {author} {\bibfnamefont {D.~A.}\ \bibnamefont {Lidar}},\ }\href
	{\doibase 10.1103/physrevlett.93.250404} {\bibfield  {journal} {\bibinfo
			{journal} {Phys. Rev. Lett.}\ }\textbf {\bibinfo {volume} {93}},\ \bibinfo
		{pages} {250404} (\bibinfo {year} {2004})}\BibitemShut {NoStop}%
	\bibitem [{\citenamefont {Refael}\ and\ \citenamefont
		{Moore}(2004)}]{Refael2004}%
	\BibitemOpen
	\bibfield  {author} {\bibinfo {author} {\bibfnamefont {G.}~\bibnamefont
			{Refael}}\ and\ \bibinfo {author} {\bibfnamefont {J.~E.}\ \bibnamefont
			{Moore}},\ }\href {\doibase 10.1103/physrevlett.93.260602} {\bibfield
		{journal} {\bibinfo  {journal} {Phys. Rev. Lett.}\ }\textbf {\bibinfo
			{volume} {93}},\ \bibinfo {pages} {260602} (\bibinfo {year}
		{2004})}\BibitemShut {NoStop}%
	\bibitem [{\citenamefont {Gu}\ \emph {et~al.}(2004)\citenamefont {Gu},
		\citenamefont {Deng}, \citenamefont {Li},\ and\ \citenamefont
		{Lin}}]{Gu2004}%
	\BibitemOpen
	\bibfield  {author} {\bibinfo {author} {\bibfnamefont {S.-J.}\ \bibnamefont
			{Gu}}, \bibinfo {author} {\bibfnamefont {S.-S.}\ \bibnamefont {Deng}},
		\bibinfo {author} {\bibfnamefont {Y.-Q.}\ \bibnamefont {Li}}, \ and\ \bibinfo
		{author} {\bibfnamefont {H.-Q.}\ \bibnamefont {Lin}},\ }\href {\doibase
		10.1103/physrevlett.93.086402} {\bibfield  {journal} {\bibinfo  {journal}
			{Phys. Rev. Lett.}\ }\textbf {\bibinfo {volume} {93}},\ \bibinfo {pages}
		{086402} (\bibinfo {year} {2004})}\BibitemShut {NoStop}%
	\bibitem [{\citenamefont {Anfossi}\ \emph {et~al.}(2005)\citenamefont
		{Anfossi}, \citenamefont {Giorda}, \citenamefont {Montorsi},\ and\
		\citenamefont {Traversa}}]{Anfossi2005}%
	\BibitemOpen
	\bibfield  {author} {\bibinfo {author} {\bibfnamefont {A.}~\bibnamefont
			{Anfossi}}, \bibinfo {author} {\bibfnamefont {P.}~\bibnamefont {Giorda}},
		\bibinfo {author} {\bibfnamefont {A.}~\bibnamefont {Montorsi}}, \ and\
		\bibinfo {author} {\bibfnamefont {F.}~\bibnamefont {Traversa}},\ }\href
	{\doibase 10.1103/physrevlett.95.056402} {\bibfield  {journal} {\bibinfo
			{journal} {Phys. Rev. Lett.}\ }\textbf {\bibinfo {volume} {95}},\ \bibinfo
		{pages} {056402} (\bibinfo {year} {2005})}\BibitemShut {NoStop}%
	\bibitem [{\citenamefont {Gu}\ \emph {et~al.}(2005)\citenamefont {Gu},
		\citenamefont {Tian},\ and\ \citenamefont {Lin}}]{Gu2005}%
	\BibitemOpen
	\bibfield  {author} {\bibinfo {author} {\bibfnamefont {S.-J.}\ \bibnamefont
			{Gu}}, \bibinfo {author} {\bibfnamefont {G.-S.}\ \bibnamefont {Tian}}, \ and\
		\bibinfo {author} {\bibfnamefont {H.-Q.}\ \bibnamefont {Lin}},\ }\href
	{\doibase 10.1103/PhysRevA.71.052322} {\bibfield  {journal} {\bibinfo
			{journal} {Phys. Rev. A}\ }\textbf {\bibinfo {volume} {71}},\ \bibinfo
		{pages} {052322} (\bibinfo {year} {2005})}\BibitemShut {NoStop}%
	\bibitem [{\citenamefont {Wei}\ \emph {et~al.}(2005)\citenamefont {Wei},
		\citenamefont {Das}, \citenamefont {Mukhopadyay}, \citenamefont
		{Vishveshwara},\ and\ \citenamefont {Goldbart}}]{Wei2005}%
	\BibitemOpen
	\bibfield  {author} {\bibinfo {author} {\bibfnamefont {T.-C.}\ \bibnamefont
			{Wei}}, \bibinfo {author} {\bibfnamefont {D.}~\bibnamefont {Das}}, \bibinfo
		{author} {\bibfnamefont {S.}~\bibnamefont {Mukhopadyay}}, \bibinfo {author}
		{\bibfnamefont {S.}~\bibnamefont {Vishveshwara}}, \ and\ \bibinfo {author}
		{\bibfnamefont {P.~M.}\ \bibnamefont {Goldbart}},\ }\href {\doibase
		10.1103/physreva.71.060305} {\bibfield  {journal} {\bibinfo  {journal} {Phys.
				Rev. A}\ }\textbf {\bibinfo {volume} {71}},\ \bibinfo {pages} {060305(R)}
		(\bibinfo {year} {2005})}\BibitemShut {NoStop}%
	\bibitem [{\citenamefont {Wu}\ \emph {et~al.}(2006)\citenamefont {Wu},
		\citenamefont {Sarandy}, \citenamefont {Lidar},\ and\ \citenamefont
		{Sham}}]{Wu2006}%
	\BibitemOpen
	\bibfield  {author} {\bibinfo {author} {\bibfnamefont {L.-A.}\ \bibnamefont
			{Wu}}, \bibinfo {author} {\bibfnamefont {M.~S.}\ \bibnamefont {Sarandy}},
		\bibinfo {author} {\bibfnamefont {D.~A.}\ \bibnamefont {Lidar}}, \ and\
		\bibinfo {author} {\bibfnamefont {L.~J.}\ \bibnamefont {Sham}},\ }\href
	{\doibase 10.1103/physreva.74.052335} {\bibfield  {journal} {\bibinfo
			{journal} {Phys. Rev. A}\ }\textbf {\bibinfo {volume} {74}},\ \bibinfo
		{pages} {052335} (\bibinfo {year} {2006})}\BibitemShut {NoStop}%
	\bibitem [{\citenamefont {de~Oliveira}\ \emph
		{et~al.}(2006{\natexlab{a}})\citenamefont {de~Oliveira}, \citenamefont
		{Rigolin}, \citenamefont {de~Oliveira},\ and\ \citenamefont
		{Miranda}}]{Oliveira2006}%
	\BibitemOpen
	\bibfield  {author} {\bibinfo {author} {\bibfnamefont {T.~R.}\ \bibnamefont
			{de~Oliveira}}, \bibinfo {author} {\bibfnamefont {G.}~\bibnamefont
			{Rigolin}}, \bibinfo {author} {\bibfnamefont {M.~C.}\ \bibnamefont
			{de~Oliveira}}, \ and\ \bibinfo {author} {\bibfnamefont {E.}~\bibnamefont
			{Miranda}},\ }\href {\doibase 10.1103/physrevlett.97.170401} {\bibfield
		{journal} {\bibinfo  {journal} {Phys. Rev. Lett.}\ }\textbf {\bibinfo
			{volume} {97}},\ \bibinfo {pages} {170401} (\bibinfo {year}
		{2006}{\natexlab{a}})}\BibitemShut {NoStop}%
	\bibitem [{\citenamefont {de~Oliveira}\ \emph
		{et~al.}(2006{\natexlab{b}})\citenamefont {de~Oliveira}, \citenamefont
		{Rigolin},\ and\ \citenamefont {de~Oliveira}}]{Oliveira2006a}%
	\BibitemOpen
	\bibfield  {author} {\bibinfo {author} {\bibfnamefont {T.~R.}\ \bibnamefont
			{de~Oliveira}}, \bibinfo {author} {\bibfnamefont {G.}~\bibnamefont
			{Rigolin}}, \ and\ \bibinfo {author} {\bibfnamefont {M.~C.}\ \bibnamefont
			{de~Oliveira}},\ }\href {\doibase 10.1103/physreva.73.010305} {\bibfield
		{journal} {\bibinfo  {journal} {Phys. Rev. A}\ }\textbf {\bibinfo {volume}
			{73}},\ \bibinfo {pages} {010305} (\bibinfo {year}
		{2006}{\natexlab{b}})}\BibitemShut {NoStop}%
	\bibitem [{\citenamefont {Buonsante}\ and\ \citenamefont
		{Vezzani}(2007)}]{Buonsante2007}%
	\BibitemOpen
	\bibfield  {author} {\bibinfo {author} {\bibfnamefont {P.}~\bibnamefont
			{Buonsante}}\ and\ \bibinfo {author} {\bibfnamefont {A.}~\bibnamefont
			{Vezzani}},\ }\href {\doibase 10.1103/physrevlett.98.110601} {\bibfield
		{journal} {\bibinfo  {journal} {Phys. Rev. Lett.}\ }\textbf {\bibinfo
			{volume} {98}},\ \bibinfo {pages} {110601} (\bibinfo {year}
		{2007})}\BibitemShut {NoStop}%
	\bibitem [{\citenamefont {Sun}\ \emph {et~al.}(2007)\citenamefont {Sun},
		\citenamefont {Wang},\ and\ \citenamefont {Sun}}]{Sun2007}%
	\BibitemOpen
	\bibfield  {author} {\bibinfo {author} {\bibfnamefont {Z.}~\bibnamefont
			{Sun}}, \bibinfo {author} {\bibfnamefont {X.}~\bibnamefont {Wang}}, \ and\
		\bibinfo {author} {\bibfnamefont {C.~P.}\ \bibnamefont {Sun}},\ }\href
	{\doibase 10.1103/physreva.75.062312} {\bibfield  {journal} {\bibinfo
			{journal} {Phys. Rev. A}\ }\textbf {\bibinfo {volume} {75}},\ \bibinfo
		{pages} {062312} (\bibinfo {year} {2007})}\BibitemShut {NoStop}%
	\bibitem [{\citenamefont {Or{\'{u}}s}(2008)}]{Orus2008}%
	\BibitemOpen
	\bibfield  {author} {\bibinfo {author} {\bibfnamefont {R.}~\bibnamefont
			{Or{\'{u}}s}},\ }\href {\doibase 10.1103/physrevlett.100.130502} {\bibfield
		{journal} {\bibinfo  {journal} {Phys. Rev. Lett.}\ }\textbf {\bibinfo
			{volume} {100}},\ \bibinfo {pages} {130502} (\bibinfo {year}
		{2008})}\BibitemShut {NoStop}%
	\bibitem [{\citenamefont {Facchi}\ \emph {et~al.}(2008)\citenamefont {Facchi},
		\citenamefont {Marzolino}, \citenamefont {Parisi}, \citenamefont {Pascazio},\
		and\ \citenamefont {Scardicchio}}]{Facchi2008}%
	\BibitemOpen
	\bibfield  {author} {\bibinfo {author} {\bibfnamefont {P.}~\bibnamefont
			{Facchi}}, \bibinfo {author} {\bibfnamefont {U.}~\bibnamefont {Marzolino}},
		\bibinfo {author} {\bibfnamefont {G.}~\bibnamefont {Parisi}}, \bibinfo
		{author} {\bibfnamefont {S.}~\bibnamefont {Pascazio}}, \ and\ \bibinfo
		{author} {\bibfnamefont {A.}~\bibnamefont {Scardicchio}},\ }\href {\doibase
		10.1103/physrevlett.101.050502} {\bibfield  {journal} {\bibinfo  {journal}
			{Phys. Rev. Lett.}\ }\textbf {\bibinfo {volume} {101}},\ \bibinfo {pages}
		{050502} (\bibinfo {year} {2008})}\BibitemShut {NoStop}%
	\bibitem [{\citenamefont {Amico}\ \emph {et~al.}(2008)\citenamefont {Amico},
		\citenamefont {Fazio}, \citenamefont {Osterloh},\ and\ \citenamefont
		{Vedral}}]{Amico2008}%
	\BibitemOpen
	\bibfield  {author} {\bibinfo {author} {\bibfnamefont {L.}~\bibnamefont
			{Amico}}, \bibinfo {author} {\bibfnamefont {R.}~\bibnamefont {Fazio}},
		\bibinfo {author} {\bibfnamefont {A.}~\bibnamefont {Osterloh}}, \ and\
		\bibinfo {author} {\bibfnamefont {V.}~\bibnamefont {Vedral}},\ }\href
	{\doibase 10.1103/RevModPhys.80.517} {\bibfield  {journal} {\bibinfo
			{journal} {Rev. Mod. Phys.}\ }\textbf {\bibinfo {volume} {80}},\ \bibinfo
		{pages} {517} (\bibinfo {year} {2008})}\BibitemShut {NoStop}%
	\bibitem [{\citenamefont {Hamma}\ \emph {et~al.}(2008)\citenamefont {Hamma},
		\citenamefont {Zhang}, \citenamefont {Haas},\ and\ \citenamefont
		{Lidar}}]{Hamma2008}%
	\BibitemOpen
	\bibfield  {author} {\bibinfo {author} {\bibfnamefont {A.}~\bibnamefont
			{Hamma}}, \bibinfo {author} {\bibfnamefont {W.}~\bibnamefont {Zhang}},
		\bibinfo {author} {\bibfnamefont {S.}~\bibnamefont {Haas}}, \ and\ \bibinfo
		{author} {\bibfnamefont {D.~A.}\ \bibnamefont {Lidar}},\ }\href {\doibase
		10.1103/physrevb.77.155111} {\bibfield  {journal} {\bibinfo  {journal} {Phys.
				Rev. B}\ }\textbf {\bibinfo {volume} {77}},\ \bibinfo {pages} {155111}
		(\bibinfo {year} {2008})}\BibitemShut {NoStop}%
	\bibitem [{\citenamefont {Rulli}\ and\ \citenamefont
		{Sarandy}(2010)}]{Rulli2010}%
	\BibitemOpen
	\bibfield  {author} {\bibinfo {author} {\bibfnamefont {C.~C.}\ \bibnamefont
			{Rulli}}\ and\ \bibinfo {author} {\bibfnamefont {M.~S.}\ \bibnamefont
			{Sarandy}},\ }\href {\doibase 10.1103/physreva.81.032334} {\bibfield
		{journal} {\bibinfo  {journal} {Phys. Rev. A}\ }\textbf {\bibinfo {volume}
			{81}},\ \bibinfo {pages} {032334} (\bibinfo {year} {2010})}\BibitemShut
	{NoStop}%
	\bibitem [{\citenamefont {Pollmann}\ \emph {et~al.}(2010)\citenamefont
		{Pollmann}, \citenamefont {Turner}, \citenamefont {Berg},\ and\ \citenamefont
		{Oshikawa}}]{Pollmann2010}%
	\BibitemOpen
	\bibfield  {author} {\bibinfo {author} {\bibfnamefont {F.}~\bibnamefont
			{Pollmann}}, \bibinfo {author} {\bibfnamefont {A.~M.}\ \bibnamefont
			{Turner}}, \bibinfo {author} {\bibfnamefont {E.}~\bibnamefont {Berg}}, \ and\
		\bibinfo {author} {\bibfnamefont {M.}~\bibnamefont {Oshikawa}},\ }\href
	{\doibase 10.1103/physrevb.81.064439} {\bibfield  {journal} {\bibinfo
			{journal} {Phys. Rev. B}\ }\textbf {\bibinfo {volume} {81}},\ \bibinfo
		{pages} {064439} (\bibinfo {year} {2010})}\BibitemShut {NoStop}%
	\bibitem [{\citenamefont {Chiara}\ \emph {et~al.}(2012)\citenamefont {Chiara},
		\citenamefont {Lepori}, \citenamefont {Lewenstein},\ and\ \citenamefont
		{Sanpera}}]{Chiara2012}%
	\BibitemOpen
	\bibfield  {author} {\bibinfo {author} {\bibfnamefont {G.~D.}\ \bibnamefont
			{Chiara}}, \bibinfo {author} {\bibfnamefont {L.}~\bibnamefont {Lepori}},
		\bibinfo {author} {\bibfnamefont {M.}~\bibnamefont {Lewenstein}}, \ and\
		\bibinfo {author} {\bibfnamefont {A.}~\bibnamefont {Sanpera}},\ }\href
	{\doibase 10.1103/physrevlett.109.237208} {\bibfield  {journal} {\bibinfo
			{journal} {Phys. Rev. Lett.}\ }\textbf {\bibinfo {volume} {109}},\ \bibinfo
		{pages} {237208} (\bibinfo {year} {2012})}\BibitemShut {NoStop}%
	\bibitem [{\citenamefont {Hofmann}\ \emph {et~al.}(2014)\citenamefont
		{Hofmann}, \citenamefont {Osterloh},\ and\ \citenamefont
		{Gühne}}]{Hofmann2014}%
	\BibitemOpen
	\bibfield  {author} {\bibinfo {author} {\bibfnamefont {M.}~\bibnamefont
			{Hofmann}}, \bibinfo {author} {\bibfnamefont {A.}~\bibnamefont {Osterloh}}, \
		and\ \bibinfo {author} {\bibfnamefont {O.}~\bibnamefont {Gühne}},\ }\href
	{\doibase 10.1103/physrevb.89.134101} {\bibfield  {journal} {\bibinfo
			{journal} {Phys. Rev. B}\ }\textbf {\bibinfo {volume} {89}},\ \bibinfo
		{pages} {134101} (\bibinfo {year} {2014})}\BibitemShut {NoStop}%
	\bibitem [{\citenamefont {Sahling}\ \emph {et~al.}(2015)\citenamefont
		{Sahling}, \citenamefont {Remenyi}, \citenamefont {Paulsen}, \citenamefont
		{Monceau}, \citenamefont {Saligrama}, \citenamefont {Marin}, \citenamefont
		{Revcolevschi}, \citenamefont {Regnault}, \citenamefont {Raymond},\ and\
		\citenamefont {Lorenzo}}]{Sahling2015}%
	\BibitemOpen
	\bibfield  {author} {\bibinfo {author} {\bibfnamefont {S.}~\bibnamefont
			{Sahling}}, \bibinfo {author} {\bibfnamefont {G.}~\bibnamefont {Remenyi}},
		\bibinfo {author} {\bibfnamefont {C.}~\bibnamefont {Paulsen}}, \bibinfo
		{author} {\bibfnamefont {P.}~\bibnamefont {Monceau}}, \bibinfo {author}
		{\bibfnamefont {V.}~\bibnamefont {Saligrama}}, \bibinfo {author}
		{\bibfnamefont {C.}~\bibnamefont {Marin}}, \bibinfo {author} {\bibfnamefont
			{A.}~\bibnamefont {Revcolevschi}}, \bibinfo {author} {\bibfnamefont {L.~P.}\
			\bibnamefont {Regnault}}, \bibinfo {author} {\bibfnamefont {S.}~\bibnamefont
			{Raymond}}, \ and\ \bibinfo {author} {\bibfnamefont {J.~E.}\ \bibnamefont
			{Lorenzo}},\ }\href {\doibase 10.1038/nphys3186} {\bibfield  {journal}
		{\bibinfo  {journal} {Nat. Phys.}\ }\textbf {\bibinfo {volume} {11}},\
		\bibinfo {pages} {255} (\bibinfo {year} {2015})}\BibitemShut {NoStop}%
	\bibitem [{\citenamefont {Bayat}(2017)}]{Bayat2017}%
	\BibitemOpen
	\bibfield  {author} {\bibinfo {author} {\bibfnamefont {A.}~\bibnamefont
			{Bayat}},\ }\href {\doibase 10.1103/physrevlett.118.036102} {\bibfield
		{journal} {\bibinfo  {journal} {Phys. Rev. Lett.}\ }\textbf {\bibinfo
			{volume} {118}},\ \bibinfo {pages} {036102} (\bibinfo {year}
		{2017})}\BibitemShut {NoStop}%
	\bibitem [{\citenamefont {Pezz{\`{e}}}\ \emph {et~al.}(2017)\citenamefont
		{Pezz{\`{e}}}, \citenamefont {Gabbrielli}, \citenamefont {Lepori},\ and\
		\citenamefont {Smerzi}}]{Pezze2017}%
	\BibitemOpen
	\bibfield  {author} {\bibinfo {author} {\bibfnamefont {L.}~\bibnamefont
			{Pezz{\`{e}}}}, \bibinfo {author} {\bibfnamefont {M.}~\bibnamefont
			{Gabbrielli}}, \bibinfo {author} {\bibfnamefont {L.}~\bibnamefont {Lepori}},
		\ and\ \bibinfo {author} {\bibfnamefont {A.}~\bibnamefont {Smerzi}},\ }\href
	{\doibase 10.1103/physrevlett.119.250401} {\bibfield  {journal} {\bibinfo
			{journal} {Phys. Rev. Lett.}\ }\textbf {\bibinfo {volume} {119}},\ \bibinfo
		{pages} {250401} (\bibinfo {year} {2017})}\BibitemShut {NoStop}%
	\bibitem [{\citenamefont {Vidmar}\ \emph {et~al.}(2018)\citenamefont {Vidmar},
		\citenamefont {Hackl}, \citenamefont {Bianchi},\ and\ \citenamefont
		{Rigol}}]{Vidmar2018}%
	\BibitemOpen
	\bibfield  {author} {\bibinfo {author} {\bibfnamefont {L.}~\bibnamefont
			{Vidmar}}, \bibinfo {author} {\bibfnamefont {L.}~\bibnamefont {Hackl}},
		\bibinfo {author} {\bibfnamefont {E.}~\bibnamefont {Bianchi}}, \ and\
		\bibinfo {author} {\bibfnamefont {M.}~\bibnamefont {Rigol}},\ }\href
	{\doibase 10.1103/physrevlett.121.220602} {\bibfield  {journal} {\bibinfo
			{journal} {Phys. Rev. Lett.}\ }\textbf {\bibinfo {volume} {121}},\ \bibinfo
		{pages} {220602} (\bibinfo {year} {2018})}\BibitemShut {NoStop}%
	\bibitem [{\citenamefont {W{\l}odzy{\'{n}}ski}\ \emph
		{et~al.}(2020)\citenamefont {W{\l}odzy{\'{n}}ski}, \citenamefont
		{P{\k{e}}cak},\ and\ \citenamefont {Sowi{\'{n}}ski}}]{Wlodzynski2020}%
	\BibitemOpen
	\bibfield  {author} {\bibinfo {author} {\bibfnamefont {D.}~\bibnamefont
			{W{\l}odzy{\'{n}}ski}}, \bibinfo {author} {\bibfnamefont {D.}~\bibnamefont
			{P{\k{e}}cak}}, \ and\ \bibinfo {author} {\bibfnamefont {T.}~\bibnamefont
			{Sowi{\'{n}}ski}},\ }\href {\doibase 10.1103/physreva.101.023604} {\bibfield
		{journal} {\bibinfo  {journal} {Phys. Rev. A}\ }\textbf {\bibinfo {volume}
			{101}},\ \bibinfo {pages} {023604} (\bibinfo {year} {2020})}\BibitemShut
	{NoStop}%
	\bibitem [{\citenamefont {Zhang}(2021)}]{Zhang2021}%
	\BibitemOpen
	\bibfield  {author} {\bibinfo {author} {\bibfnamefont {J.}~\bibnamefont
			{Zhang}},\ }\href {\doibase 10.1103/physrevb.104.205112} {\bibfield
		{journal} {\bibinfo  {journal} {Phys. Rev. B}\ }\textbf {\bibinfo {volume}
			{104}},\ \bibinfo {pages} {205112} (\bibinfo {year} {2021})}\BibitemShut
	{NoStop}%
	\bibitem [{\citenamefont {Yi}\ \emph {et~al.}(2006)\citenamefont {Yi},
		\citenamefont {Cui},\ and\ \citenamefont {Wang}}]{Yi2006}%
	\BibitemOpen
	\bibfield  {author} {\bibinfo {author} {\bibfnamefont {X.~X.}\ \bibnamefont
			{Yi}}, \bibinfo {author} {\bibfnamefont {H.~T.}\ \bibnamefont {Cui}}, \ and\
		\bibinfo {author} {\bibfnamefont {L.~C.}\ \bibnamefont {Wang}},\ }\href
	{\doibase 10.1103/physreva.74.054102} {\bibfield  {journal} {\bibinfo
			{journal} {Phys. Rev. A}\ }\textbf {\bibinfo {volume} {74}},\ \bibinfo
		{pages} {054102} (\bibinfo {year} {2006})}\BibitemShut {NoStop}%
	\bibitem [{\citenamefont {Dillenschneider}(2008)}]{Dillenschneider2008}%
	\BibitemOpen
	\bibfield  {author} {\bibinfo {author} {\bibfnamefont {R.}~\bibnamefont
			{Dillenschneider}},\ }\href {\doibase 10.1103/physrevb.78.224413} {\bibfield
		{journal} {\bibinfo  {journal} {Phys. Rev. B}\ }\textbf {\bibinfo {volume}
			{78}},\ \bibinfo {pages} {224413} (\bibinfo {year} {2008})}\BibitemShut
	{NoStop}%
	\bibitem [{\citenamefont {Sarandy}(2009)}]{Sarandy2009}%
	\BibitemOpen
	\bibfield  {author} {\bibinfo {author} {\bibfnamefont {M.~S.}\ \bibnamefont
			{Sarandy}},\ }\href {\doibase 10.1103/physreva.80.022108} {\bibfield
		{journal} {\bibinfo  {journal} {Phys. Rev. A}\ }\textbf {\bibinfo {volume}
			{80}},\ \bibinfo {pages} {022108} (\bibinfo {year} {2009})}\BibitemShut
	{NoStop}%
	\bibitem [{\citenamefont {Cui}\ \emph {et~al.}(2010)\citenamefont {Cui},
		\citenamefont {Cao},\ and\ \citenamefont {Fan}}]{Cui2010}%
	\BibitemOpen
	\bibfield  {author} {\bibinfo {author} {\bibfnamefont {J.}~\bibnamefont
			{Cui}}, \bibinfo {author} {\bibfnamefont {J.-P.}\ \bibnamefont {Cao}}, \ and\
		\bibinfo {author} {\bibfnamefont {H.}~\bibnamefont {Fan}},\ }\href {\doibase
		10.1103/physreva.82.022319} {\bibfield  {journal} {\bibinfo  {journal} {Phys.
				Rev. A}\ }\textbf {\bibinfo {volume} {82}},\ \bibinfo {pages} {022319}
		(\bibinfo {year} {2010})}\BibitemShut {NoStop}%
	\bibitem [{\citenamefont {Sun}\ \emph {et~al.}(2010)\citenamefont {Sun},
		\citenamefont {Li}, \citenamefont {Yao}, \citenamefont {Du}, \citenamefont
		{Liu}, \citenamefont {Luo}, \citenamefont {Li},\ and\ \citenamefont
		{Li}}]{Sun2010}%
	\BibitemOpen
	\bibfield  {author} {\bibinfo {author} {\bibfnamefont {Z.-Y.}\ \bibnamefont
			{Sun}}, \bibinfo {author} {\bibfnamefont {L.}~\bibnamefont {Li}}, \bibinfo
		{author} {\bibfnamefont {K.-L.}\ \bibnamefont {Yao}}, \bibinfo {author}
		{\bibfnamefont {G.-H.}\ \bibnamefont {Du}}, \bibinfo {author} {\bibfnamefont
			{J.-W.}\ \bibnamefont {Liu}}, \bibinfo {author} {\bibfnamefont
			{B.}~\bibnamefont {Luo}}, \bibinfo {author} {\bibfnamefont {N.}~\bibnamefont
			{Li}}, \ and\ \bibinfo {author} {\bibfnamefont {H.-N.}\ \bibnamefont {Li}},\
	}\href {\doibase 10.1103/physreva.82.032310} {\bibfield  {journal} {\bibinfo
			{journal} {Phys. Rev. A}\ }\textbf {\bibinfo {volume} {82}},\ \bibinfo
		{pages} {032310} (\bibinfo {year} {2010})}\BibitemShut {NoStop}%
	\bibitem [{\citenamefont {Ren}\ \emph {et~al.}(2012)\citenamefont {Ren},
		\citenamefont {Wu},\ and\ \citenamefont {Zhu}}]{2012Ren60305}%
	\BibitemOpen
	\bibfield  {author} {\bibinfo {author} {\bibfnamefont {J.}~\bibnamefont
			{Ren}}, \bibinfo {author} {\bibfnamefont {Y.-Z.}\ \bibnamefont {Wu}}, \ and\
		\bibinfo {author} {\bibfnamefont {S.-Q.}\ \bibnamefont {Zhu}},\ }\href
	{\doibase 10.1088/0256-307x/29/6/060305} {\bibfield  {journal} {\bibinfo
			{journal} {Chin. Phys. Lett.}\ }\textbf {\bibinfo {volume} {29}},\ \bibinfo
		{pages} {060305} (\bibinfo {year} {2012})}\BibitemShut {NoStop}%
	\bibitem [{\citenamefont {Ye}\ \emph {et~al.}(2017)\citenamefont {Ye},
		\citenamefont {Li}, \citenamefont {Zhao}, \citenamefont {Zhang},\ and\
		\citenamefont {Fei}}]{2017Ye}%
	\BibitemOpen
	\bibfield  {author} {\bibinfo {author} {\bibfnamefont {B.-L.}\ \bibnamefont
			{Ye}}, \bibinfo {author} {\bibfnamefont {B.}~\bibnamefont {Li}}, \bibinfo
		{author} {\bibfnamefont {L.-J.}\ \bibnamefont {Zhao}}, \bibinfo {author}
		{\bibfnamefont {H.-J.}\ \bibnamefont {Zhang}}, \ and\ \bibinfo {author}
		{\bibfnamefont {S.-M.}\ \bibnamefont {Fei}},\ }\href {\doibase
		10.1007/s11433-016-0425-x} {\bibfield  {journal} {\bibinfo  {journal} {Sci.
				China Phys. Mech. Astron.}\ }\textbf {\bibinfo {volume} {60}},\ \bibinfo
		{pages} {030311} (\bibinfo {year} {2017})}\BibitemShut {NoStop}%
	\bibitem [{\citenamefont {Du}\ \emph {et~al.}(2021)\citenamefont {Du},
		\citenamefont {Zhang}, \citenamefont {Zhou},\ and\ \citenamefont
		{Tong}}]{2021Du12418}%
	\BibitemOpen
	\bibfield  {author} {\bibinfo {author} {\bibfnamefont {M.-M.}\ \bibnamefont
			{Du}}, \bibinfo {author} {\bibfnamefont {D.-J.}\ \bibnamefont {Zhang}},
		\bibinfo {author} {\bibfnamefont {Z.-Y.}\ \bibnamefont {Zhou}}, \ and\
		\bibinfo {author} {\bibfnamefont {D.~M.}\ \bibnamefont {Tong}},\ }\href
	{\doibase 10.1103/PhysRevA.104.012418} {\bibfield  {journal} {\bibinfo
			{journal} {Phys. Rev. A}\ }\textbf {\bibinfo {volume} {104}},\ \bibinfo
		{pages} {012418} (\bibinfo {year} {2021})}\BibitemShut {NoStop}%
	\bibitem [{\citenamefont {Ollivier}\ and\ \citenamefont
		{Zurek}(2001)}]{Ollivier2001}%
	\BibitemOpen
	\bibfield  {author} {\bibinfo {author} {\bibfnamefont {H.}~\bibnamefont
			{Ollivier}}\ and\ \bibinfo {author} {\bibfnamefont {W.~H.}\ \bibnamefont
			{Zurek}},\ }\href {\doibase 10.1103/physrevlett.88.017901} {\bibfield
		{journal} {\bibinfo  {journal} {Phys. Rev. Lett.}\ }\textbf {\bibinfo
			{volume} {88}},\ \bibinfo {pages} {017901} (\bibinfo {year}
		{2001})}\BibitemShut {NoStop}%
	\bibitem [{\citenamefont {Deng}\ \emph {et~al.}(2012)\citenamefont {Deng},
		\citenamefont {Wu}, \citenamefont {Chen}, \citenamefont {Gu}, \citenamefont
		{Yu},\ and\ \citenamefont {Oh}}]{Deng2012}%
	\BibitemOpen
	\bibfield  {author} {\bibinfo {author} {\bibfnamefont {D.-L.}\ \bibnamefont
			{Deng}}, \bibinfo {author} {\bibfnamefont {C.}~\bibnamefont {Wu}}, \bibinfo
		{author} {\bibfnamefont {J.-L.}\ \bibnamefont {Chen}}, \bibinfo {author}
		{\bibfnamefont {S.-J.}\ \bibnamefont {Gu}}, \bibinfo {author} {\bibfnamefont
			{S.}~\bibnamefont {Yu}}, \ and\ \bibinfo {author} {\bibfnamefont {C.~H.}\
			\bibnamefont {Oh}},\ }\href {\doibase 10.1103/physreva.86.032305} {\bibfield
		{journal} {\bibinfo  {journal} {Phys. Rev. A}\ }\textbf {\bibinfo {volume}
			{86}},\ \bibinfo {pages} {032305} (\bibinfo {year} {2012})}\BibitemShut
	{NoStop}%
	\bibitem [{\citenamefont {Justino}\ and\ \citenamefont
		{de~Oliveira}(2012)}]{Justino2012}%
	\BibitemOpen
	\bibfield  {author} {\bibinfo {author} {\bibfnamefont {L.}~\bibnamefont
			{Justino}}\ and\ \bibinfo {author} {\bibfnamefont {T.~R.}\ \bibnamefont
			{de~Oliveira}},\ }\href {\doibase 10.1103/physreva.85.052128} {\bibfield
		{journal} {\bibinfo  {journal} {Phys. Rev. A}\ }\textbf {\bibinfo {volume}
			{85}},\ \bibinfo {pages} {052128} (\bibinfo {year} {2012})}\BibitemShut
	{NoStop}%
	\bibitem [{\citenamefont {Groisman}\ \emph {et~al.}(2005)\citenamefont
		{Groisman}, \citenamefont {Popescu},\ and\ \citenamefont
		{Winter}}]{Groisman2005}%
	\BibitemOpen
	\bibfield  {author} {\bibinfo {author} {\bibfnamefont {B.}~\bibnamefont
			{Groisman}}, \bibinfo {author} {\bibfnamefont {S.}~\bibnamefont {Popescu}}, \
		and\ \bibinfo {author} {\bibfnamefont {A.}~\bibnamefont {Winter}},\ }\href
	{\doibase 10.1103/physreva.72.032317} {\bibfield  {journal} {\bibinfo
			{journal} {Phys. Rev. A}\ }\textbf {\bibinfo {volume} {72}},\ \bibinfo
		{pages} {032317} (\bibinfo {year} {2005})}\BibitemShut {NoStop}%
	\bibitem [{\citenamefont {Baumgratz}\ \emph {et~al.}(2014)\citenamefont
		{Baumgratz}, \citenamefont {Cramer},\ and\ \citenamefont
		{Plenio}}]{Baumgratz2014}%
	\BibitemOpen
	\bibfield  {author} {\bibinfo {author} {\bibfnamefont {T.}~\bibnamefont
			{Baumgratz}}, \bibinfo {author} {\bibfnamefont {M.}~\bibnamefont {Cramer}}, \
		and\ \bibinfo {author} {\bibfnamefont {M.}~\bibnamefont {Plenio}},\ }\href
	{\doibase 10.1103/physrevlett.113.140401} {\bibfield  {journal} {\bibinfo
			{journal} {Phys. Rev. Lett.}\ }\textbf {\bibinfo {volume} {113}},\ \bibinfo
		{pages} {140401} (\bibinfo {year} {2014})}\BibitemShut {NoStop}%
	\bibitem [{\citenamefont {Tan}\ \emph {et~al.}(2016)\citenamefont {Tan},
		\citenamefont {Kwon}, \citenamefont {Park},\ and\ \citenamefont
		{Jeong}}]{Tan2016}%
	\BibitemOpen
	\bibfield  {author} {\bibinfo {author} {\bibfnamefont {K.~C.}\ \bibnamefont
			{Tan}}, \bibinfo {author} {\bibfnamefont {H.}~\bibnamefont {Kwon}}, \bibinfo
		{author} {\bibfnamefont {C.-Y.}\ \bibnamefont {Park}}, \ and\ \bibinfo
		{author} {\bibfnamefont {H.}~\bibnamefont {Jeong}},\ }\href {\doibase
		10.1103/physreva.94.022329} {\bibfield  {journal} {\bibinfo  {journal} {Phys.
				Rev. A}\ }\textbf {\bibinfo {volume} {94}},\ \bibinfo {pages} {022329}
		(\bibinfo {year} {2016})}\BibitemShut {NoStop}%
	\bibitem [{\citenamefont {Yu}\ \emph {et~al.}(2016{\natexlab{a}})\citenamefont
		{Yu}, \citenamefont {Zhang}, \citenamefont {Xu},\ and\ \citenamefont
		{Tong}}]{2016Yu60302}%
	\BibitemOpen
	\bibfield  {author} {\bibinfo {author} {\bibfnamefont {X.-D.}\ \bibnamefont
			{Yu}}, \bibinfo {author} {\bibfnamefont {D.-J.}\ \bibnamefont {Zhang}},
		\bibinfo {author} {\bibfnamefont {G.~F.}\ \bibnamefont {Xu}}, \ and\ \bibinfo
		{author} {\bibfnamefont {D.~M.}\ \bibnamefont {Tong}},\ }\href {\doibase
		10.1103/PhysRevA.94.060302} {\bibfield  {journal} {\bibinfo  {journal} {Phys.
				Rev. A}\ }\textbf {\bibinfo {volume} {94}},\ \bibinfo {pages} {060302(R)}
		(\bibinfo {year} {2016}{\natexlab{a}})}\BibitemShut {NoStop}%
	\bibitem [{\citenamefont {Yu}\ \emph {et~al.}(2016{\natexlab{b}})\citenamefont
		{Yu}, \citenamefont {Zhang}, \citenamefont {Liu},\ and\ \citenamefont
		{Tong}}]{2016Yu60303}%
	\BibitemOpen
	\bibfield  {author} {\bibinfo {author} {\bibfnamefont {X.-D.}\ \bibnamefont
			{Yu}}, \bibinfo {author} {\bibfnamefont {D.-J.}\ \bibnamefont {Zhang}},
		\bibinfo {author} {\bibfnamefont {C.~L.}\ \bibnamefont {Liu}}, \ and\
		\bibinfo {author} {\bibfnamefont {D.~M.}\ \bibnamefont {Tong}},\ }\href
	{\doibase 10.1103/PhysRevA.93.060303} {\bibfield  {journal} {\bibinfo
			{journal} {Phys. Rev. A}\ }\textbf {\bibinfo {volume} {93}},\ \bibinfo
		{pages} {060303(R)} (\bibinfo {year} {2016}{\natexlab{b}})}\BibitemShut
	{NoStop}%
	\bibitem [{\citenamefont {Ma}\ \emph {et~al.}(2016)\citenamefont {Ma},
		\citenamefont {Zhao}, \citenamefont {Fei},\ and\ \citenamefont
		{Long}}]{Ma2016a}%
	\BibitemOpen
	\bibfield  {author} {\bibinfo {author} {\bibfnamefont {T.}~\bibnamefont
			{Ma}}, \bibinfo {author} {\bibfnamefont {M.-J.}\ \bibnamefont {Zhao}},
		\bibinfo {author} {\bibfnamefont {S.-M.}\ \bibnamefont {Fei}}, \ and\
		\bibinfo {author} {\bibfnamefont {G.-L.}\ \bibnamefont {Long}},\ }\href
	{\doibase 10.1103/physreva.94.042312} {\bibfield  {journal} {\bibinfo
			{journal} {Phys. Rev. A}\ }\textbf {\bibinfo {volume} {94}},\ \bibinfo
		{pages} {042312} (\bibinfo {year} {2016})}\BibitemShut {NoStop}%
	\bibitem [{\citenamefont {Ma}\ \emph {et~al.}(2017)\citenamefont {Ma},
		\citenamefont {Zhao}, \citenamefont {Zhang}, \citenamefont {Fei},\ and\
		\citenamefont {Long}}]{Ma2017}%
	\BibitemOpen
	\bibfield  {author} {\bibinfo {author} {\bibfnamefont {T.}~\bibnamefont
			{Ma}}, \bibinfo {author} {\bibfnamefont {M.-J.}\ \bibnamefont {Zhao}},
		\bibinfo {author} {\bibfnamefont {H.-J.}\ \bibnamefont {Zhang}}, \bibinfo
		{author} {\bibfnamefont {S.-M.}\ \bibnamefont {Fei}}, \ and\ \bibinfo
		{author} {\bibfnamefont {G.-L.}\ \bibnamefont {Long}},\ }\href {\doibase
		10.1103/physreva.95.042328} {\bibfield  {journal} {\bibinfo  {journal} {Phys.
				Rev. A}\ }\textbf {\bibinfo {volume} {95}},\ \bibinfo {pages} {042328}
		(\bibinfo {year} {2017})}\BibitemShut {NoStop}%
	\bibitem [{\citenamefont {Zhang}\ \emph {et~al.}(2018)\citenamefont {Zhang},
		\citenamefont {Liu}, \citenamefont {Yu},\ and\ \citenamefont
		{Tong}}]{2018Zhang170501}%
	\BibitemOpen
	\bibfield  {author} {\bibinfo {author} {\bibfnamefont {D.-J.}\ \bibnamefont
			{Zhang}}, \bibinfo {author} {\bibfnamefont {C.~L.}\ \bibnamefont {Liu}},
		\bibinfo {author} {\bibfnamefont {X.-D.}\ \bibnamefont {Yu}}, \ and\ \bibinfo
		{author} {\bibfnamefont {D.~M.}\ \bibnamefont {Tong}},\ }\href {\doibase
		10.1103/PhysRevLett.120.170501} {\bibfield  {journal} {\bibinfo  {journal}
			{Phys. Rev. Lett.}\ }\textbf {\bibinfo {volume} {120}},\ \bibinfo {pages}
		{170501} (\bibinfo {year} {2018})}\BibitemShut {NoStop}%
	\bibitem [{\citenamefont {Liu}\ and\ \citenamefont
		{Zhou}(2019)}]{2019Liu70402}%
	\BibitemOpen
	\bibfield  {author} {\bibinfo {author} {\bibfnamefont {C.~L.}\ \bibnamefont
			{Liu}}\ and\ \bibinfo {author} {\bibfnamefont {D.~L.}\ \bibnamefont {Zhou}},\
	}\href {\doibase 10.1103/PhysRevLett.123.070402} {\bibfield  {journal}
		{\bibinfo  {journal} {Phys. Rev. Lett.}\ }\textbf {\bibinfo {volume} {123}},\
		\bibinfo {pages} {070402} (\bibinfo {year} {2019})}\BibitemShut {NoStop}%
	\bibitem [{\citenamefont {Jin}\ \emph {et~al.}(2021)\citenamefont {Jin},
		\citenamefont {Yang}, \citenamefont {Fei}, \citenamefont {Li-Jost},
		\citenamefont {Wang}, \citenamefont {Long},\ and\ \citenamefont
		{Qiao}}]{2021Jin280311}%
	\BibitemOpen
	\bibfield  {author} {\bibinfo {author} {\bibfnamefont {Z.-X.}\ \bibnamefont
			{Jin}}, \bibinfo {author} {\bibfnamefont {L.-M.}\ \bibnamefont {Yang}},
		\bibinfo {author} {\bibfnamefont {S.-M.}\ \bibnamefont {Fei}}, \bibinfo
		{author} {\bibfnamefont {X.}~\bibnamefont {Li-Jost}}, \bibinfo {author}
		{\bibfnamefont {Z.-X.}\ \bibnamefont {Wang}}, \bibinfo {author}
		{\bibfnamefont {G.-L.}\ \bibnamefont {Long}}, \ and\ \bibinfo {author}
		{\bibfnamefont {C.-F.}\ \bibnamefont {Qiao}},\ }\href {\doibase
		10.1007/s11433-021-1709-9} {\bibfield  {journal} {\bibinfo  {journal} {Sci.
				China Phys. Mech. Astron.}\ }\textbf {\bibinfo {volume} {64}},\ \bibinfo
		{pages} {280311} (\bibinfo {year} {2021})}\BibitemShut {NoStop}%
	\bibitem [{\citenamefont {Chen}\ \emph {et~al.}(2016)\citenamefont {Chen},
		\citenamefont {Cui}, \citenamefont {Zhang},\ and\ \citenamefont
		{Fan}}]{Chen2016}%
	\BibitemOpen
	\bibfield  {author} {\bibinfo {author} {\bibfnamefont {J.-J.}\ \bibnamefont
			{Chen}}, \bibinfo {author} {\bibfnamefont {J.}~\bibnamefont {Cui}}, \bibinfo
		{author} {\bibfnamefont {Y.-R.}\ \bibnamefont {Zhang}}, \ and\ \bibinfo
		{author} {\bibfnamefont {H.}~\bibnamefont {Fan}},\ }\href {\doibase
		10.1103/physreva.94.022112} {\bibfield  {journal} {\bibinfo  {journal} {Phys.
				Rev. A}\ }\textbf {\bibinfo {volume} {94}},\ \bibinfo {pages} {022112}
		(\bibinfo {year} {2016})}\BibitemShut {NoStop}%
	\bibitem [{\citenamefont {Qin}\ \emph {et~al.}(2018)\citenamefont {Qin},
		\citenamefont {Ren},\ and\ \citenamefont {Zhang}}]{Qin2018}%
	\BibitemOpen
	\bibfield  {author} {\bibinfo {author} {\bibfnamefont {M.}~\bibnamefont
			{Qin}}, \bibinfo {author} {\bibfnamefont {Z.}~\bibnamefont {Ren}}, \ and\
		\bibinfo {author} {\bibfnamefont {X.}~\bibnamefont {Zhang}},\ }\href
	{\doibase 10.1103/physreva.98.012303} {\bibfield  {journal} {\bibinfo
			{journal} {Phys. Rev. A}\ }\textbf {\bibinfo {volume} {98}},\ \bibinfo
		{pages} {012303} (\bibinfo {year} {2018})}\BibitemShut {NoStop}%
	\bibitem [{\citenamefont {Hu}\ \emph {et~al.}(2020)\citenamefont {Hu},
		\citenamefont {Gao},\ and\ \citenamefont {Fan}}]{Hu2020}%
	\BibitemOpen
	\bibfield  {author} {\bibinfo {author} {\bibfnamefont {M.-L.}\ \bibnamefont
			{Hu}}, \bibinfo {author} {\bibfnamefont {Y.-Y.}\ \bibnamefont {Gao}}, \ and\
		\bibinfo {author} {\bibfnamefont {H.}~\bibnamefont {Fan}},\ }\href {\doibase
		10.1103/physreva.101.032305} {\bibfield  {journal} {\bibinfo  {journal}
			{Phys. Rev. A}\ }\textbf {\bibinfo {volume} {101}},\ \bibinfo {pages}
		{032305} (\bibinfo {year} {2020})}\BibitemShut {NoStop}%
	\bibitem [{\citenamefont {Hu}\ \emph {et~al.}(2021)\citenamefont {Hu},
		\citenamefont {Fang},\ and\ \citenamefont {Fan}}]{2021Hu}%
	\BibitemOpen
	\bibfield  {author} {\bibinfo {author} {\bibfnamefont {M.-L.}\ \bibnamefont
			{Hu}}, \bibinfo {author} {\bibfnamefont {F.}~\bibnamefont {Fang}}, \ and\
		\bibinfo {author} {\bibfnamefont {H.}~\bibnamefont {Fan}},\ }\href {\doibase
		10.1103/PhysRevA.104.062416} {\bibfield  {journal} {\bibinfo  {journal}
			{Phys. Rev. A}\ }\textbf {\bibinfo {volume} {104}},\ \bibinfo {pages}
		{062416} (\bibinfo {year} {2021})}\BibitemShut {NoStop}%
	\bibitem [{\citenamefont {Radhakrishnan}\ \emph {et~al.}(2017)\citenamefont
		{Radhakrishnan}, \citenamefont {Ermakov},\ and\ \citenamefont
		{Byrnes}}]{Radhakrishnan2017}%
	\BibitemOpen
	\bibfield  {author} {\bibinfo {author} {\bibfnamefont {C.}~\bibnamefont
			{Radhakrishnan}}, \bibinfo {author} {\bibfnamefont {I.}~\bibnamefont
			{Ermakov}}, \ and\ \bibinfo {author} {\bibfnamefont {T.}~\bibnamefont
			{Byrnes}},\ }\href {\doibase 10.1103/physreva.96.012341} {\bibfield
		{journal} {\bibinfo  {journal} {Phys. Rev. A}\ }\textbf {\bibinfo {volume}
			{96}},\ \bibinfo {pages} {012341} (\bibinfo {year} {2017})}\BibitemShut
	{NoStop}%
	\bibitem [{\citenamefont {Mondal}\ \emph {et~al.}(2017)\citenamefont {Mondal},
		\citenamefont {Pramanik},\ and\ \citenamefont {Pati}}]{Mondal2017}%
	\BibitemOpen
	\bibfield  {author} {\bibinfo {author} {\bibfnamefont {D.}~\bibnamefont
			{Mondal}}, \bibinfo {author} {\bibfnamefont {T.}~\bibnamefont {Pramanik}}, \
		and\ \bibinfo {author} {\bibfnamefont {A.~K.}\ \bibnamefont {Pati}},\ }\href
	{\doibase 10.1103/physreva.95.010301} {\bibfield  {journal} {\bibinfo
			{journal} {Phys. Rev. A}\ }\textbf {\bibinfo {volume} {95}},\ \bibinfo
		{pages} {010301(R)} (\bibinfo {year} {2017})}\BibitemShut {NoStop}%
	\bibitem [{\citenamefont {Tan}\ and\ \citenamefont {Jeong}(2018)}]{Tan2018}%
	\BibitemOpen
	\bibfield  {author} {\bibinfo {author} {\bibfnamefont {K.~C.}\ \bibnamefont
			{Tan}}\ and\ \bibinfo {author} {\bibfnamefont {H.}~\bibnamefont {Jeong}},\
	}\href {\doibase 10.1103/physrevlett.121.220401} {\bibfield  {journal}
		{\bibinfo  {journal} {Phys. Rev. Lett.}\ }\textbf {\bibinfo {volume} {121}},\
		\bibinfo {pages} {220401} (\bibinfo {year} {2018})}\BibitemShut {NoStop}%
	\bibitem [{\citenamefont {Li}\ and\ \citenamefont {Sun}(2018)}]{Li2018}%
	\BibitemOpen
	\bibfield  {author} {\bibinfo {author} {\bibfnamefont {S.-P.}\ \bibnamefont
			{Li}}\ and\ \bibinfo {author} {\bibfnamefont {Z.-H.}\ \bibnamefont {Sun}},\
	}\href {\doibase 10.1103/physreva.98.022317} {\bibfield  {journal} {\bibinfo
			{journal} {Phys. Rev. A}\ }\textbf {\bibinfo {volume} {98}},\ \bibinfo
		{pages} {022317} (\bibinfo {year} {2018})}\BibitemShut {NoStop}%
	\bibitem [{\citenamefont {Xie}(2020)}]{Xie2020}%
	\BibitemOpen
	\bibfield  {author} {\bibinfo {author} {\bibfnamefont {Y.-X.}\ \bibnamefont
			{Xie}},\ }\href {\doibase 10.1002/pssb.202000322} {\bibfield  {journal}
		{\bibinfo  {journal} {Phys. Status Solidi B}\ }\textbf {\bibinfo {volume}
			{258}},\ \bibinfo {pages} {2000322} (\bibinfo {year} {2020})}\BibitemShut
	{NoStop}%
	\bibitem [{\citenamefont {Xie}\ and\ \citenamefont {Zhang}(2020)}]{Xie2020a}%
	\BibitemOpen
	\bibfield  {author} {\bibinfo {author} {\bibfnamefont {Y.-X.}\ \bibnamefont
			{Xie}}\ and\ \bibinfo {author} {\bibfnamefont {Y.-H.}\ \bibnamefont
			{Zhang}},\ }\href {\doibase 10.1088/1612-202x/ab6aa4} {\bibfield  {journal}
		{\bibinfo  {journal} {Laser Phys. Lett.}\ }\textbf {\bibinfo {volume} {17}},\
		\bibinfo {pages} {035206} (\bibinfo {year} {2020})}\BibitemShut {NoStop}%
	\bibitem [{\citenamefont {Ye}\ and\ \citenamefont {Zhang}(2020)}]{Ye2020a}%
	\BibitemOpen
	\bibfield  {author} {\bibinfo {author} {\bibfnamefont {B.}~\bibnamefont
			{Ye}}\ and\ \bibinfo {author} {\bibfnamefont {Z.}~\bibnamefont {Zhang}},\
	}\href {\doibase 10.1142/s0217732321500024} {\bibfield  {journal} {\bibinfo
			{journal} {Mod. Phys. Lett. A}\ }\textbf {\bibinfo {volume} {36}},\ \bibinfo
		{pages} {2150002} (\bibinfo {year} {2020})}\BibitemShut {NoStop}%
	\bibitem [{\citenamefont {Ye}\ \emph {et~al.}(2018)\citenamefont {Ye},
		\citenamefont {Li}, \citenamefont {Wang}, \citenamefont {Li-Jost},\ and\
		\citenamefont {Fei}}]{Ye2018c}%
	\BibitemOpen
	\bibfield  {author} {\bibinfo {author} {\bibfnamefont {B.-L.}\ \bibnamefont
			{Ye}}, \bibinfo {author} {\bibfnamefont {B.}~\bibnamefont {Li}}, \bibinfo
		{author} {\bibfnamefont {Z.-X.}\ \bibnamefont {Wang}}, \bibinfo {author}
		{\bibfnamefont {X.}~\bibnamefont {Li-Jost}}, \ and\ \bibinfo {author}
		{\bibfnamefont {S.-M.}\ \bibnamefont {Fei}},\ }\href {\doibase
		10.1007/s11433-018-9262-9} {\bibfield  {journal} {\bibinfo  {journal} {Sci.
				China Phys. Mech. Astron.}\ }\textbf {\bibinfo {volume} {61}},\ \bibinfo
		{pages} {110312} (\bibinfo {year} {2018})}\BibitemShut {NoStop}%
	\bibitem [{\citenamefont {Wang}\ \emph {et~al.}(2010)\citenamefont {Wang},
		\citenamefont {Ma}, \citenamefont {Gu},\ and\ \citenamefont
		{Lin}}]{Wang2010}%
	\BibitemOpen
	\bibfield  {author} {\bibinfo {author} {\bibfnamefont {Z.}~\bibnamefont
			{Wang}}, \bibinfo {author} {\bibfnamefont {T.}~\bibnamefont {Ma}}, \bibinfo
		{author} {\bibfnamefont {S.-J.}\ \bibnamefont {Gu}}, \ and\ \bibinfo {author}
		{\bibfnamefont {H.-Q.}\ \bibnamefont {Lin}},\ }\href {\doibase
		10.1103/physreva.81.062350} {\bibfield  {journal} {\bibinfo  {journal} {Phys.
				Rev. A}\ }\textbf {\bibinfo {volume} {81}},\ \bibinfo {pages} {062350}
		(\bibinfo {year} {2010})}\BibitemShut {NoStop}%
	\bibitem [{\citenamefont {Yang}\ \emph {et~al.}(2008)\citenamefont {Yang},
		\citenamefont {Gu}, \citenamefont {Sun},\ and\ \citenamefont
		{Lin}}]{Yang2008}%
	\BibitemOpen
	\bibfield  {author} {\bibinfo {author} {\bibfnamefont {S.}~\bibnamefont
			{Yang}}, \bibinfo {author} {\bibfnamefont {S.-J.}\ \bibnamefont {Gu}},
		\bibinfo {author} {\bibfnamefont {C.-P.}\ \bibnamefont {Sun}}, \ and\
		\bibinfo {author} {\bibfnamefont {H.-Q.}\ \bibnamefont {Lin}},\ }\href
	{\doibase 10.1103/physreva.78.012304} {\bibfield  {journal} {\bibinfo
			{journal} {Phys. Rev. A}\ }\textbf {\bibinfo {volume} {78}},\ \bibinfo
		{pages} {012304} (\bibinfo {year} {2008})}\BibitemShut {NoStop}%
	\bibitem [{\citenamefont {Malvezzi}\ \emph {et~al.}(2016)\citenamefont
		{Malvezzi}, \citenamefont {Karpat}, \citenamefont {{\c{C}}akmak},
		\citenamefont {Fanchini}, \citenamefont {Debarba},\ and\ \citenamefont
		{Vianna}}]{Malvezzi2016}%
	\BibitemOpen
	\bibfield  {author} {\bibinfo {author} {\bibfnamefont {A.~L.}\ \bibnamefont
			{Malvezzi}}, \bibinfo {author} {\bibfnamefont {G.}~\bibnamefont {Karpat}},
		\bibinfo {author} {\bibfnamefont {B.}~\bibnamefont {{\c{C}}akmak}}, \bibinfo
		{author} {\bibfnamefont {F.~F.}\ \bibnamefont {Fanchini}}, \bibinfo {author}
		{\bibfnamefont {T.}~\bibnamefont {Debarba}}, \ and\ \bibinfo {author}
		{\bibfnamefont {R.~O.}\ \bibnamefont {Vianna}},\ }\href {\doibase
		10.1103/physrevb.93.184428} {\bibfield  {journal} {\bibinfo  {journal} {Phys.
				Rev. B}\ }\textbf {\bibinfo {volume} {93}},\ \bibinfo {pages} {184428}
		(\bibinfo {year} {2016})}\BibitemShut {NoStop}%
	\bibitem [{\citenamefont {Ye}\ \emph {et~al.}(2020)\citenamefont {Ye},
		\citenamefont {Xue}, \citenamefont {Fang}, \citenamefont {Liu}, \citenamefont
		{Wu}, \citenamefont {Zhou},\ and\ \citenamefont {Yang}}]{Ye2020}%
	\BibitemOpen
	\bibfield  {author} {\bibinfo {author} {\bibfnamefont {B.-L.}\ \bibnamefont
			{Ye}}, \bibinfo {author} {\bibfnamefont {L.-Y.}\ \bibnamefont {Xue}},
		\bibinfo {author} {\bibfnamefont {Y.-L.}\ \bibnamefont {Fang}}, \bibinfo
		{author} {\bibfnamefont {S.}~\bibnamefont {Liu}}, \bibinfo {author}
		{\bibfnamefont {Q.-C.}\ \bibnamefont {Wu}}, \bibinfo {author} {\bibfnamefont
			{Y.-H.}\ \bibnamefont {Zhou}}, \ and\ \bibinfo {author} {\bibfnamefont
			{C.-P.}\ \bibnamefont {Yang}},\ }\href {\doibase 10.1016/j.physe.2019.113690}
	{\bibfield  {journal} {\bibinfo  {journal} {Phys. E}\ }\textbf {\bibinfo
			{volume} {115}},\ \bibinfo {pages} {113690} (\bibinfo {year}
		{2020})}\BibitemShut {NoStop}%
	\bibitem [{\citenamefont {Mandel}\ and\ \citenamefont
		{Wolf}(1995)}]{Mandel1995}%
	\BibitemOpen
	\bibfield  {author} {\bibinfo {author} {\bibfnamefont {L.}~\bibnamefont
			{Mandel}}\ and\ \bibinfo {author} {\bibfnamefont {E.}~\bibnamefont {Wolf}},\
	}\href@noop {} {\emph {\bibinfo {title} {Optical Coherence and Quantum
				Optics}}}\ (\bibinfo  {publisher} {Cambridge University Press, Cambridge},\
	\bibinfo {year} {1995})\BibitemShut {NoStop}%
	\bibitem [{\citenamefont {Patoary}\ \emph {et~al.}(2019)\citenamefont
		{Patoary}, \citenamefont {Kulkarni},\ and\ \citenamefont
		{Jha}}]{Patoary2019}%
	\BibitemOpen
	\bibfield  {author} {\bibinfo {author} {\bibfnamefont {A.~S.~M.}\
			\bibnamefont {Patoary}}, \bibinfo {author} {\bibfnamefont {G.}~\bibnamefont
			{Kulkarni}}, \ and\ \bibinfo {author} {\bibfnamefont {A.~K.}\ \bibnamefont
			{Jha}},\ }\href {\doibase 10.1364/josab.36.002765} {\bibfield  {journal}
		{\bibinfo  {journal} {J. Opt. Soc. Am. B}\ }\textbf {\bibinfo {volume}
			{36}},\ \bibinfo {pages} {2765} (\bibinfo {year} {2019})}\BibitemShut
	{NoStop}%
	\bibitem [{\citenamefont {Zhang}\ \emph
		{et~al.}(2016{\natexlab{a}})\citenamefont {Zhang}, \citenamefont {Huang},\
		and\ \citenamefont {Tong}}]{2016Zhang12117}%
	\BibitemOpen
	\bibfield  {author} {\bibinfo {author} {\bibfnamefont {D.-J.}\ \bibnamefont
			{Zhang}}, \bibinfo {author} {\bibfnamefont {H.-L.}\ \bibnamefont {Huang}}, \
		and\ \bibinfo {author} {\bibfnamefont {D.~M.}\ \bibnamefont {Tong}},\ }\href
	{\doibase 10.1103/physreva.93.012117} {\bibfield  {journal} {\bibinfo
			{journal} {Phys. Rev. A}\ }\textbf {\bibinfo {volume} {93}},\ \bibinfo
		{pages} {012117} (\bibinfo {year} {2016}{\natexlab{a}})}\BibitemShut
	{NoStop}%
	\bibitem [{\citenamefont {Zhang}\ \emph
		{et~al.}(2016{\natexlab{b}})\citenamefont {Zhang}, \citenamefont {Yu},
		\citenamefont {Huang},\ and\ \citenamefont {Tong}}]{2016Zhang52132}%
	\BibitemOpen
	\bibfield  {author} {\bibinfo {author} {\bibfnamefont {D.-J.}\ \bibnamefont
			{Zhang}}, \bibinfo {author} {\bibfnamefont {X.-D.}\ \bibnamefont {Yu}},
		\bibinfo {author} {\bibfnamefont {H.-L.}\ \bibnamefont {Huang}}, \ and\
		\bibinfo {author} {\bibfnamefont {D.~M.}\ \bibnamefont {Tong}},\ }\href
	{\doibase 10.1103/PhysRevA.94.052132} {\bibfield  {journal} {\bibinfo
			{journal} {Phys. Rev. A}\ }\textbf {\bibinfo {volume} {94}},\ \bibinfo
		{pages} {052132} (\bibinfo {year} {2016}{\natexlab{b}})}\BibitemShut
	{NoStop}%
	\bibitem [{\citenamefont {{D.-J. Zhang and Q.-h. Wang and J.
				Gong}}(2019{\natexlab{a}})}]{2019Zhang42104}%
	\BibitemOpen
	\bibfield  {author} {\bibinfo {author} {\bibnamefont {{D.-J. Zhang and Q.-h.
					Wang and J. Gong}}},\ }\href {\doibase 10.1103/PhysRevA.99.042104} {\bibfield
		{journal} {\bibinfo  {journal} {Phys. Rev. A}\ }\textbf {\bibinfo {volume}
			{99}},\ \bibinfo {pages} {042104} (\bibinfo {year}
		{2019}{\natexlab{a}})}\BibitemShut {NoStop}%
	\bibitem [{\citenamefont {Zhang}\ and\ \citenamefont
		{Tong}(2022)}]{2022Zhang81}%
	\BibitemOpen
	\bibfield  {author} {\bibinfo {author} {\bibfnamefont {D.-J.}\ \bibnamefont
			{Zhang}}\ and\ \bibinfo {author} {\bibfnamefont {D.~M.}\ \bibnamefont
			{Tong}},\ }\href {\doibase 10.1038/s41534-022-00588-2} {\bibfield  {journal}
		{\bibinfo  {journal} {npj Quant. Inf.}\ }\textbf {\bibinfo {volume} {8}},\
		\bibinfo {pages} {81} (\bibinfo {year} {2022})}\BibitemShut {NoStop}%
	\bibitem [{\citenamefont {{D.-J. Zhang and Q.-h. Wang and J.
				Gong}}(2019{\natexlab{b}})}]{2019ZWG62121}%
	\BibitemOpen
	\bibfield  {author} {\bibinfo {author} {\bibnamefont {{D.-J. Zhang and Q.-h.
					Wang and J. Gong}}},\ }\href {\doibase 10.1103/PhysRevA.100.062121}
	{\bibfield  {journal} {\bibinfo  {journal} {Phys. Rev. A}\ }\textbf {\bibinfo
			{volume} {100}},\ \bibinfo {pages} {062121} (\bibinfo {year}
		{2019}{\natexlab{b}})}\BibitemShut {NoStop}%
	\bibitem [{\citenamefont {Zhang}\ and\ \citenamefont
		{Gong}(2020)}]{2020Zhang23418}%
	\BibitemOpen
	\bibfield  {author} {\bibinfo {author} {\bibfnamefont {D.-J.}\ \bibnamefont
			{Zhang}}\ and\ \bibinfo {author} {\bibfnamefont {J.}~\bibnamefont {Gong}},\
	}\href {\doibase 10.1103/PhysRevResearch.2.023418} {\bibfield  {journal}
		{\bibinfo  {journal} {Phys. Rev. Research}\ }\textbf {\bibinfo {volume}
			{2}},\ \bibinfo {pages} {023418} (\bibinfo {year} {2020})}\BibitemShut
	{NoStop}%
	\bibitem [{\citenamefont {Kagalwala}\ \emph {et~al.}(2012)\citenamefont
		{Kagalwala}, \citenamefont {Giuseppe}, \citenamefont {Abouraddy},\ and\
		\citenamefont {Saleh}}]{Kagalwala2012}%
	\BibitemOpen
	\bibfield  {author} {\bibinfo {author} {\bibfnamefont {K.~H.}\ \bibnamefont
			{Kagalwala}}, \bibinfo {author} {\bibfnamefont {G.~D.}\ \bibnamefont
			{Giuseppe}}, \bibinfo {author} {\bibfnamefont {A.~F.}\ \bibnamefont
			{Abouraddy}}, \ and\ \bibinfo {author} {\bibfnamefont {B.~E.~A.}\
			\bibnamefont {Saleh}},\ }\href {\doibase 10.1038/nphoton.2012.312} {\bibfield
		{journal} {\bibinfo  {journal} {Nat. Photon.}\ }\textbf {\bibinfo {volume}
			{7}},\ \bibinfo {pages} {72} (\bibinfo {year} {2012})}\BibitemShut {NoStop}%
	\bibitem [{\citenamefont {Svozil{\'{\i}}k}\ \emph {et~al.}(2015)\citenamefont
		{Svozil{\'{\i}}k}, \citenamefont {Vall{\'{e}}s}, \citenamefont
		{Pe{\v{r}}ina},\ and\ \citenamefont {Torres}}]{Svozilik2015}%
	\BibitemOpen
	\bibfield  {author} {\bibinfo {author} {\bibfnamefont {J.}~\bibnamefont
			{Svozil{\'{\i}}k}}, \bibinfo {author} {\bibfnamefont {A.}~\bibnamefont
			{Vall{\'{e}}s}}, \bibinfo {author} {\bibfnamefont {J.}~\bibnamefont
			{Pe{\v{r}}ina}}, \ and\ \bibinfo {author} {\bibfnamefont {J.~P.}\
			\bibnamefont {Torres}},\ }\href {\doibase 10.1103/physrevlett.115.220501}
	{\bibfield  {journal} {\bibinfo  {journal} {Phys. Rev. Lett.}\ }\textbf
		{\bibinfo {volume} {115}},\ \bibinfo {pages} {220501} (\bibinfo {year}
		{2015})}\BibitemShut {NoStop}%
	\bibitem [{\citenamefont {Yao}\ \emph {et~al.}(2016)\citenamefont {Yao},
		\citenamefont {Dong}, \citenamefont {Xiao},\ and\ \citenamefont
		{Sun}}]{Yao2016}%
	\BibitemOpen
	\bibfield  {author} {\bibinfo {author} {\bibfnamefont {Y.}~\bibnamefont
			{Yao}}, \bibinfo {author} {\bibfnamefont {G.~H.}\ \bibnamefont {Dong}},
		\bibinfo {author} {\bibfnamefont {X.}~\bibnamefont {Xiao}}, \ and\ \bibinfo
		{author} {\bibfnamefont {C.~P.}\ \bibnamefont {Sun}},\ }\href {\doibase
		10.1038/srep32010} {\bibfield  {journal} {\bibinfo  {journal} {Sci. Rep}\
		}\textbf {\bibinfo {volume} {6}},\ \bibinfo {pages} {32010} (\bibinfo {year}
		{2016})}\BibitemShut {NoStop}%
	\bibitem [{\citenamefont {{\v{C}}ernoch}\ \emph {et~al.}(2018)\citenamefont
		{{\v{C}}ernoch}, \citenamefont {Bartkiewicz}, \citenamefont {Lemr},\ and\
		\citenamefont {Soubusta}}]{Cernoch2018}%
	\BibitemOpen
	\bibfield  {author} {\bibinfo {author} {\bibfnamefont {A.}~\bibnamefont
			{{\v{C}}ernoch}}, \bibinfo {author} {\bibfnamefont {K.}~\bibnamefont
			{Bartkiewicz}}, \bibinfo {author} {\bibfnamefont {K.}~\bibnamefont {Lemr}}, \
		and\ \bibinfo {author} {\bibfnamefont {J.}~\bibnamefont {Soubusta}},\ }\href
	{\doibase 10.1103/physreva.97.042305} {\bibfield  {journal} {\bibinfo
			{journal} {Phys. Rev. A}\ }\textbf {\bibinfo {volume} {97}},\ \bibinfo
		{pages} {042305} (\bibinfo {year} {2018})}\BibitemShut {NoStop}%
	\bibitem [{\citenamefont {Kalaga}\ \emph {et~al.}(2018)\citenamefont {Kalaga},
		\citenamefont {Leo{\'{n}}ski},\ and\ \citenamefont
		{Pe{\v{r}}ina}}]{Kalaga2018}%
	\BibitemOpen
	\bibfield  {author} {\bibinfo {author} {\bibfnamefont {J.~K.}\ \bibnamefont
			{Kalaga}}, \bibinfo {author} {\bibfnamefont {W.}~\bibnamefont
			{Leo{\'{n}}ski}}, \ and\ \bibinfo {author} {\bibfnamefont {J.}~\bibnamefont
			{Pe{\v{r}}ina}},\ }\href {\doibase 10.1103/physreva.97.042110} {\bibfield
		{journal} {\bibinfo  {journal} {Phys. Rev. A}\ }\textbf {\bibinfo {volume}
			{97}},\ \bibinfo {pages} {042110} (\bibinfo {year} {2018})}\BibitemShut
	{NoStop}%
	\bibitem [{\citenamefont {Fan}\ \emph {et~al.}(2019)\citenamefont {Fan},
		\citenamefont {Sun}, \citenamefont {Ding}, \citenamefont {Ming},
		\citenamefont {Yang}, \citenamefont {Wang},\ and\ \citenamefont
		{Ye}}]{Fan2019}%
	\BibitemOpen
	\bibfield  {author} {\bibinfo {author} {\bibfnamefont {X.-G.}\ \bibnamefont
			{Fan}}, \bibinfo {author} {\bibfnamefont {W.-Y.}\ \bibnamefont {Sun}},
		\bibinfo {author} {\bibfnamefont {Z.-Y.}\ \bibnamefont {Ding}}, \bibinfo
		{author} {\bibfnamefont {F.}~\bibnamefont {Ming}}, \bibinfo {author}
		{\bibfnamefont {H.}~\bibnamefont {Yang}}, \bibinfo {author} {\bibfnamefont
			{D.}~\bibnamefont {Wang}}, \ and\ \bibinfo {author} {\bibfnamefont
			{L.}~\bibnamefont {Ye}},\ }\href {\doibase 10.1088/1367-2630/ab41b1}
	{\bibfield  {journal} {\bibinfo  {journal} {New J. Phys.}\ }\textbf {\bibinfo
			{volume} {21}},\ \bibinfo {pages} {093053} (\bibinfo {year}
		{2019})}\BibitemShut {NoStop}%
	\bibitem [{\citenamefont {Du}\ and\ \citenamefont {Tong}(2021)}]{Du2021}%
	\BibitemOpen
	\bibfield  {author} {\bibinfo {author} {\bibfnamefont {M.-M.}\ \bibnamefont
			{Du}}\ and\ \bibinfo {author} {\bibfnamefont {D.~M.}\ \bibnamefont {Tong}},\
	}\href {\doibase 10.1103/physreva.103.032407} {\bibfield  {journal} {\bibinfo
			{journal} {Phys. Rev. A}\ }\textbf {\bibinfo {volume} {103}},\ \bibinfo
		{pages} {032407} (\bibinfo {year} {2021})}\BibitemShut {NoStop}%
	\bibitem [{\citenamefont {Brukner}\ and\ \citenamefont
		{Zeilinger}(1999)}]{Brukner1999}%
	\BibitemOpen
	\bibfield  {author} {\bibinfo {author} {\bibfnamefont {{\v{C}}.}~\bibnamefont
			{Brukner}}\ and\ \bibinfo {author} {\bibfnamefont {A.}~\bibnamefont
			{Zeilinger}},\ }\href {\doibase 10.1103/physrevlett.83.3354} {\bibfield
		{journal} {\bibinfo  {journal} {Phys. Rev. Lett.}\ }\textbf {\bibinfo
			{volume} {83}},\ \bibinfo {pages} {3354} (\bibinfo {year}
		{1999})}\BibitemShut {NoStop}%
	\bibitem [{\citenamefont {Kimura}(2003)}]{2003Kimura339}%
	\BibitemOpen
	\bibfield  {author} {\bibinfo {author} {\bibfnamefont {G.}~\bibnamefont
			{Kimura}},\ }\href {\doibase 10.1016/S0375-9601(03)00941-1} {\bibfield
		{journal} {\bibinfo  {journal} {Phys. Lett. A}\ }\textbf {\bibinfo {volume}
			{314}},\ \bibinfo {pages} {339} (\bibinfo {year} {2003})}\BibitemShut
	{NoStop}%
	\bibitem [{\citenamefont {Designolle}\ \emph {et~al.}(2021)\citenamefont
		{Designolle}, \citenamefont {Uola}, \citenamefont {Luoma},\ and\
		\citenamefont {Brunner}}]{2021Designolle220404}%
	\BibitemOpen
	\bibfield  {author} {\bibinfo {author} {\bibfnamefont {S.}~\bibnamefont
			{Designolle}}, \bibinfo {author} {\bibfnamefont {R.}~\bibnamefont {Uola}},
		\bibinfo {author} {\bibfnamefont {K.}~\bibnamefont {Luoma}}, \ and\ \bibinfo
		{author} {\bibfnamefont {N.}~\bibnamefont {Brunner}},\ }\href {\doibase
		10.1103/PhysRevLett.126.220404} {\bibfield  {journal} {\bibinfo  {journal}
			{Phys. Rev. Lett.}\ }\textbf {\bibinfo {volume} {126}},\ \bibinfo {pages}
		{220404} (\bibinfo {year} {2021})}\BibitemShut {NoStop}%
	\bibitem [{\citenamefont {Li}\ \emph {et~al.}(2020)\citenamefont {Li},
		\citenamefont {Fei}, \citenamefont {Xiong},\ and\ \citenamefont
		{Wang}}]{2020Li280312}%
	\BibitemOpen
	\bibfield  {author} {\bibinfo {author} {\bibfnamefont {M.-S.}\ \bibnamefont
			{Li}}, \bibinfo {author} {\bibfnamefont {S.-M.}\ \bibnamefont {Fei}},
		\bibinfo {author} {\bibfnamefont {Z.-X.}\ \bibnamefont {Xiong}}, \ and\
		\bibinfo {author} {\bibfnamefont {Y.-L.}\ \bibnamefont {Wang}},\ }\href
	{\doibase 10.1007/s11433-020-1562-4} {\bibfield  {journal} {\bibinfo
			{journal} {Sci. China Phys. Mech. Astron.}\ }\textbf {\bibinfo {volume}
			{63}},\ \bibinfo {pages} {280312} (\bibinfo {year} {2020})}\BibitemShut
	{NoStop}%
	\bibitem [{\citenamefont {Shiroishi}(2005)}]{Shiroishi2005}%
	\BibitemOpen
	\bibfield  {author} {\bibinfo {author} {\bibfnamefont {M.}~\bibnamefont
			{Shiroishi}},\ }\href {\doibase https://doi.org/10.1143/JPSJS.74S.47}
	{\bibfield  {journal} {\bibinfo  {journal} {J. Phys. Soc. Jpn.}\ }\textbf
		{\bibinfo {volume} {74}},\ \bibinfo {pages} {47} (\bibinfo {year}
		{2005})}\BibitemShut {NoStop}%
	\bibitem [{\citenamefont {Yang}\ and\ \citenamefont
		{Yang}(1966{\natexlab{a}})}]{Yang1966}%
	\BibitemOpen
	\bibfield  {author} {\bibinfo {author} {\bibfnamefont {C.~N.}\ \bibnamefont
			{Yang}}\ and\ \bibinfo {author} {\bibfnamefont {C.~P.}\ \bibnamefont
			{Yang}},\ }\href {\doibase 10.1103/physrev.150.321} {\bibfield  {journal}
		{\bibinfo  {journal} {Phys. Rev.}\ }\textbf {\bibinfo {volume} {150}},\
		\bibinfo {pages} {321} (\bibinfo {year} {1966}{\natexlab{a}})}\BibitemShut
	{NoStop}%
	\bibitem [{\citenamefont {Yang}\ and\ \citenamefont
		{Yang}(1966{\natexlab{b}})}]{Yang1966a}%
	\BibitemOpen
	\bibfield  {author} {\bibinfo {author} {\bibfnamefont {C.~N.}\ \bibnamefont
			{Yang}}\ and\ \bibinfo {author} {\bibfnamefont {C.~P.}\ \bibnamefont
			{Yang}},\ }\href {\doibase 10.1103/physrev.150.327} {\bibfield  {journal}
		{\bibinfo  {journal} {Phys. Rev.}\ }\textbf {\bibinfo {volume} {150}},\
		\bibinfo {pages} {327} (\bibinfo {year} {1966}{\natexlab{b}})}\BibitemShut
	{NoStop}%
	\bibitem [{\citenamefont {Yang}(2005)}]{2005Yang30302}%
	\BibitemOpen
	\bibfield  {author} {\bibinfo {author} {\bibfnamefont {M.-F.}\ \bibnamefont
			{Yang}},\ }\href {\doibase 10.1103/PhysRevA.71.030302} {\bibfield  {journal}
		{\bibinfo  {journal} {Phys. Rev. A}\ }\textbf {\bibinfo {volume} {71}},\
		\bibinfo {pages} {030302(R)} (\bibinfo {year} {2005})}\BibitemShut {NoStop}%
	\bibitem [{\citenamefont {Kitaev}(2006)}]{Kitaev2006}%
	\BibitemOpen
	\bibfield  {author} {\bibinfo {author} {\bibfnamefont {A.}~\bibnamefont
			{Kitaev}},\ }\href {\doibase 10.1016/j.aop.2005.10.005} {\bibfield  {journal}
		{\bibinfo  {journal} {Ann. Phys.}\ }\textbf {\bibinfo {volume} {321}},\
		\bibinfo {pages} {2} (\bibinfo {year} {2006})}\BibitemShut {NoStop}%
	\bibitem [{\citenamefont {Zhang}\ and\ \citenamefont
		{Gong}(2018)}]{2018Zhang52101}%
	\BibitemOpen
	\bibfield  {author} {\bibinfo {author} {\bibfnamefont {D.-J.}\ \bibnamefont
			{Zhang}}\ and\ \bibinfo {author} {\bibfnamefont {J.}~\bibnamefont {Gong}},\
	}\href {\doibase 10.1103/physreva.98.052101} {\bibfield  {journal} {\bibinfo
			{journal} {Phys. Rev. A}\ }\textbf {\bibinfo {volume} {98}},\ \bibinfo
		{pages} {052101} (\bibinfo {year} {2018})}\BibitemShut {NoStop}%
\end{thebibliography}
%

\end{document}